\documentclass[reprint,aps,final,longbibliography,superscriptaddress,nobibnotes,footinbib,amsfonts,showpacs,letterpaper,jmp,amssymb,pdftex]{revtex4-1}
\usepackage{graphicx,multirow,bm,amsmath}
\usepackage{booktabs}
\usepackage{hyperref}
\usepackage{bbm}
\usepackage[usenames,dvipsnames]{color}
\usepackage[table,xcdraw]{xcolor}

\newcommand{\lbnl}{Energy Technologies Area, Lawrence Berkeley National Laboratory, Berkeley, CA 94720, USA}
\newcommand{\northwestern}{Department of Materials Science \& Engineering, Northwestern University, Evanston, IL 60208, USA}
\newcommand{\yale}{Department of Applied Physics, Yale University, New Haven, CT 06511, USA}
\newcommand{\esi}{Energy Sciences Institute, Yale University, West Haven, CT 06516, USA}

\begin{document}

	\title{High Thermoelectric Performance and Defect Energetics of Multipocketed Full-Heusler Compounds}
	\author{Junsoo Park}
	\email{qkwnstn@gmail.com}
	\affiliation{\lbnl}
	\author{Yi Xia}
	\affiliation{\northwestern}
	\author{Alex M. Ganose}
	\affiliation{\lbnl}
	\author{Anubhav Jain}
	\email{ajain@lbl.gov}
	\affiliation{\lbnl} 
	\author{Vidvuds Ozoli\c{n}\v{s}}
	\email{vidvuds.ozolins@yale.edu}
	\affiliation{\yale}
	\affiliation{\esi} 
	\date{\today} 
	\begin{abstract}

We report first-principles density-functional study of electron-phonon interactions and thermoelectric transport properties of full-Heusler compounds Sr$_{2}$BiAu and Sr$_{2}$SbAu. Our results show that ultrahigh intrinsic bulk thermoelectric performance across a wide range of temperatures is physically possible and point to the presence of multiply degenerate and highly dispersive carrier pockets as the key factor for achieving it. Sr$_{2}$BiAu, which features ten energy-aligned low effective mass pockets (six along $\Gamma-X$ and four at $L$), is predicted to deliver $n$-type $zT=0.4-4.9$ at $T=100-700$~K. Comparison with the previously investigated Ba$_{2}$BiAu compound shows that the additional $L$-pockets in Sr$_{2}$BiAu significantly increase its low-temperature power factor to a maximum value of $12$~mW~m$^{-1}$~K$^{-2}$ near $T=300$~K.  However, at high temperatures the power factor of Sr$_{2}$BiAu drops below that of Ba$_{2}$BiAu because the $L$ states are heavier and subject to strong scattering by phonon deformation as opposed to the lighter $\Gamma-X$ states that are limited by polar-optical scattering. Sr$_{2}$SbAu is predicted to deliver lower $n$-type of $zT=3.4$ at $T=750$~K due to appreciable misalignment between the $L$ and $\Gamma-X$ carrier pockets, generally heavier scattering, and slightly higher lattice thermal conductivity. Soft acoustic modes, responsible for low lattice thermal conductivity, also increase vibrational entropies and high-temperature stability of the Heusler compounds, suggesting that their experimental synthesis may be feasible. The dominant intrinsic defects are found to be Au vacancies, which drive the Fermi level towards the conduction band and work in favor of $n$-doping.
	\end{abstract}
	\maketitle

\section{Introduction}

Thermoelectricity is a clean energy harvesting technology that allows direct interconversion between heat and electric current. The indicator of thermoelectric efficiency is the dimensionless figure of merit known as $zT$. To date, bulk thermoelectric materials have not overcome $zT=3$, with $zT=2$ only achieved in a few chalcogenide compounds \cite{cu2seuher,alincuse,snsescience1,snsescience2,snsenatcomm,last,pbtes,gesbinte,geteferroelectric,getejoule}. Materials that deliver commercially-relevant performance below room temperature are particularly scarce, where alloys of Bi$_{2}$Te$_{3}$ \cite{bisbtenano,bisbtedislocation,bitese} and Mg$_{3}$Sb$_{2}$-Mg$_{3}$Bi$_{2}$ \cite{mg3sb2discovery,lowtmg3sb2,mg3sb2revelation,mg3bi2cooling} are essentially the only materials with $zT$ near 1. This is unfortunate since many industrial applications, including refrigeration and spacecraft propulsion, would greatly benefit from efficient thermoelectrics at room-to-cryogenic temperatures \cite{thermoelectricsystem,advancesinthermoelectricmaterials,birdeye,spaceauto}.

The dearth of efficient thermoelectrics especially at low temperatures can easily be inferred from the definition $zT= \frac{\alpha^{2}\sigma T}{\kappa}$. Here, $\alpha^{2}\sigma$ is the thermoelectric power factor (PF), composed of the Seebeck coefficient ($\alpha$) and electronic conductivity ($\sigma$). The total thermal conductivity ($\kappa$) is the sum of lattice thermal conductivity and electronic thermal conductivity ($\kappa_{\text{lat}}+\kappa_{\text{e}}$). A desirable thermoelectric material requires high $\alpha$ and $\sigma$ with low $\kappa$. Unfortunately, such a combination is inherently difficult to achieve \cite{complex,newandold,intuition,thermoelectricmaterials,compromisesynergy,advancesinthermoelectrics}. At low temperatures, due to small $T$ and high $\kappa_{\text{lat}}$, designing for high PF is all the more indispensable, but unfortunately high PF is generally limited by the counterproductive behaviors of $\alpha$ and $\sigma$. High $\sigma$ must then arise from high mobility( $\mu$) since attempts to boost it via doping necessarily suppresses $\alpha$. A large number of band pockets is generally thought to enhance $\sigma$ at presumably little to no expense in $\alpha$ because it can deliver higher carrier concentration for given Fermi level. These considerations are best represented by bands of 1) small effective mass ($m$) capable of producing $\mu$ and 2) high band degeneracy or pocket multiplicity \cite{loweffmass,compromisesynergy,computationalthermoelectrics,halfheuslerbanddegeneracy,complexityfactor,materialdescriptors}. 

As a culmination of these concepts, full-Heusler Ba$_{2}$BiAu ($n$-type) was recently studied based on a rigorous treatment of electron-phonon and phonon-phonon scattering, and has led to the prediction of an unprecedentedly high $zT\approx5$ at 700 K and a promising $zT\approx1.5$ at 300 K \cite{ba2biau,ba2biauredo}. In this compound, one highly dispersive conduction band pocket along sixfold degenerate $\Gamma-X$ proved critical to the high PF. Meanwhile, $\kappa_{\text{lat}}$ is minimal due to anharmonic rattling of Au atoms, a trait shared by this class of full-Heusler compounds \cite{ultralowheusler}. The study showcased a rare coexistance of very high PF and ultralow $\kappa_{\text{lat}}$ for bulk thermoelectrics --- albeit without consideration of dopability and the experimental realizability of the compound.

In the present work, we achieve the following. 1) We show that analogous but multi-pocketed full-Heusler compounds, in particular Sr$_{2}$BiAu, can theoretically attain even higher thermoelectric performance across a broader temperature spectrum, which is especially niche at low temperatures. 2) We analyze the benefit of pocket multiplicity in the form of accidental (non-symmetry-related) degeneracy, which is conditional upon the similarities of the pockets. 3) We predict that the Heusler phase stability is entropically favored and that the lowest-energy defect, Au vacancy, is favorable for $n$-doping.

\section{Computational Methods}

\subsection{Electronic Structures}
The electronic structure is calculated with Quantum Espresso \cite{qespresso1,qespresso2} with the Optimized Norm-Conserving Vanderbilt pseudopotentials \cite{oncv1,oncv2,oncv3} and Perdew-Burke-Ernzerhof (PBE) exchange-correlation functional \cite{pbe} with and without spin-orbit coupling (SOC) for comparisons. Plane-wave cutoff of 100 Ry is used. In order to obtain more accurate band gaps, advanced functionals such as the modified Becke-Johnson potential (mBJ) by Tran and Blaha \cite{mbj} and the Heyd-Scuseria-Ernzerhof hybrid-exchange functional (HSE06) \cite{hse1,hse2} were used.

\subsection{Electron-phonon Scattering}
In treating electron-phonon scattering, we first compute electronic states and e-ph interaction matrix elements at a coarse $8\times8\times8$ \textbf{k}-point mesh, using phonon perturbations computed at a coarse $4\times4\times4$ \textbf{q}-point mesh using density functional perturbation theory (DFPT) \cite{ponceprb,poncejcp}. Then with the EPW package \cite{epw1,epw2,epw3,epwreview} we interpolate electronic states, phonons, and the matrix elements onto dense $40\times40\times40$ \textbf{k}-point and \textbf{q}-point meshes through maximally localized Wannier functions \cite{mlwfcomposite,mlwfentangled,wannier90}. Long-ranged polar optical scattering matrix elements are added on the dense \textbf{k}-mesh \cite{epwpolar}. The imaginary part of the resulting electron self-energy leads directly to band-and-$\textbf{k}$-dependent electron lifetimes ($\tau_{\nu\mathbf{k}}$) limited by electron-phonon (e-ph) scattering. Further theoretical details can be found in Supplemental Material at [URL will be inserted by publisher] \cite{supplementary}. 

\subsection{Electron Transport}
With $\tau_{\nu\mathbf{k}}$ as inputs, we employ the Boltzmann transport formalism (implemented in BoltzTraP \cite{boltztrap} modified in-house) in the relaxation time approximation (RTA) to compute electron transport properties:
\begin{equation}\label{eq:sigma}
\sigma=\frac{1}{\Omega N_{\mathbf{k}}} \sum_{\nu\mathbf{k}} (\tau v^{2})_{\nu\mathbf{k}}\left(-\frac{\partial f}{\partial E}\right)_{\nu\mathbf{k}},
\end{equation}
\begin{equation}\label{eq:alpha}
\alpha=\frac{\sigma^{-1}}{\Omega TN_{\mathbf{k}}} \sum_{\nu\mathbf{k}} (\tau v^{2})_{\nu\mathbf{k}}(E_{\text{F}}-E_{\nu\mathbf{k}})\left(-\frac{\partial f}{\partial E}\right)_{\nu\mathbf{k}},
\end{equation}
\begin{equation}\label{eq:kappae}
\kappa_{e}=\frac{1}{\Omega TN_{\mathbf{k}}} \sum_{\nu\mathbf{k}} (\tau v^{2})_{\nu\mathbf{k}}(E_{\text{F}}-E_{\nu\mathbf{k}})^{2}\left(-\frac{\partial f}{\partial E}\right)_{\nu\mathbf{k}}-\alpha^{2}\sigma T.
\end{equation}
The validity of RTA coupled with e-ph matrix elements calculated via DFPT and Wannier interpolation has been well-established by multiple recent instances of application that approximated experimental measurements well \cite{pbteepw1,pbteepw2,nbfesbnatcomm,cosiyi}. In performing Eqs. \ref{eq:sigma}--\ref{eq:kappae} we utilize the band structure calculated with SOC and the band gap value from mBJ+SOC for consistent comparison with our previous study on Ba$_{2}$BiAu. For Sr$_{2}$BiAu, since the effect of SOC on the electronic structure or phonon is minimal (as will be shown), SOC is neglected for the computations of electron-phonon scattering, due to the computational expense. For Sr$_{2}$SbAu, however, SOC substantially impacts the electronic structure and is therefore included in electron-phonon scattering computations. 


\subsection{Stability and Defects}

We use Vienna \textit{Ab initio} Simulations Package (VASP) \cite{vasp1,vasp2,vasp3,vasp4} throughout this section to perform DFT total energy calculations for both competing phases and defective supercells. We also incorporate SOC and use the projector-augmented wave (PAW) pseudopotentials \cite{paw} with the PBE functional throughout.

All binary and ternary phases that could potentially form from the compositions of the Heuslers compounds available on Materials Project \cite{materialsproject} and Inorganic Crystal Structure Database \cite{icsd1,icsd2,icsd3} are considered for the evaluation of phase stability. Sr$_{2}$BiAu has thirteen competing binary and ternary phases, while Ba$_{2}$BiAu has ten, Sr$_{2}$SbAu twelve, and Ba$_{2}$SbAu fifteen. We construct the ternary phase diagrams using the calculated formation energies (shown in Fig. S4 of Supplemental Material \cite{supplementary}), which reveal the phase fields under each of which various pairs of competing phases may coexist with the compounds of interest. The equilibrium chemical potentials are derived using the corresponding phase fields. Of note, all compounds with energies within the numerical noise of DFT ($\sim$10 meV per atom) from the convex hull were placed on the hull.

We consider all possible vacancy and antisite intrinsic point defects, and employ the standard supercell approach. We create host and defective supercells that are $2\times2\times2$ expansions of the fully relaxed conventional cubic unit cells of the compounds. The host supercells contain 128 atoms, of which 64 are Ba/Sr atoms, 32 are Bi/Sb atoms, and 32 are Au atoms. This is allowed for all types of point defects because a full-Heusler crystal structure remains identical upon the exchange of lattice sites between the Ba/Sr atoms and the Bi/Sb and Au atoms. For charged supercells, electrons are either removed or added according to the charge. For total energies of defective supercells, a plane-wave cut-off energy of 600 eV and a $2\times2\times2$ \textbf{k}-point mesh are used throughout all self-consistent calculations of defective supercells. All relaxations are performed with the Methfessel-Paxton's smearing scheme \cite{mpmethod} to properly treat the metallic characteristics of the cells with charged defects.

Point defect formation energies are calculated as
\begin{equation}\label{eq:defectef}
\Delta E_{f}^{\text{D}}=E^{\text{D}}-E^{\text{Host}}-\sum_{a}\Delta N_{a}\mu_{a}+q(E_{v}+E_{\text{F}})+E_{\text{cor}},
\end{equation}
where $E^{\text{D}}$ is the total energy of a defective cell, $E^{\text{Host}}$ is the total energy of a host full-Heusler cell into which a defect is introduced, $q$ is the charge on the defect and $\mu_{a}$ is the chemical potential of element $a$ in the compounds (e.g., as determined by the phase diagram). $\Delta N_{a}$ is the excess (positive) or deficient (negative) number of atoms of element $a$ in the defective cell relative to the host cell. For instance, if the defect is  Bi$_{\text{Au}}$ antisite (Bi in place of Au), then $\Delta N_{\text{Bi}}=1$ and $\Delta N_{\text{Au}}=-1$. $E_{\text{F}}$ is a free parameter and represents the Fermi level as counted positively up from $E_{v}$, which is the energy required to remove an electron from a given host, i.e., the valence band maximum (VBM) of the host compound. Lastly, $E_{\text{cor}}$ is a correction term for finite-sized supercells, which experience several fictitious effects. 

Charged defects experience fictitious electrostatic interactions between periodic images of the defect, due to periodic boundary condition, and interactions between the defect and the homogeneous, jellium-like background charge that enforces overall charge-neutrality. These are corrected by the method of Makov and Payne \cite{makovpayne},
\begin{equation}\label{eq:makovpayne}
E_{\text{cor}}=\frac{q^{2}\gamma}{2\epsilon L}-\frac{2\pi qQ}{3\epsilon L^{3}},
\end{equation}
where $\gamma$ is the Madelung constant, $Q$ is the quadrupole moment, $L$ is the supercell lattice parameter, and $\epsilon$ is the dielectric constant of the host compound. While more sophisticated correction schemes have been proposed \cite{supercellsizeprl,supercellsizepss,lanyzunger}, we do not employ them because 1) the compounds have high dielectric constants, 2) cell sizes used are large enough for Eq. \ref{eq:makovpayne} to be acceptable ($L>16$ \r{A}), and 3) other methods will likely not change the main conclusions we draw. Band-gap correction also must be performed to reference the defect energies to more realistic band edges of the host compounds, for which we use HSE06 with SOC. This treatment scheme of choice is based on HSE06's credible track record of preserving the band-edge-relative defect energies calculated with PBE when aligned to a common reference level (achieved with the local electrostatic potential) \cite{defecthse1,defecthse2,defecthse3,defecthse4,defecthse5}. Potential adjustment as well as band-filling corrections are made \cite{lanyzunger}. These corrections are computed using the aide code (not yet published). Additional details on calculations of phase stability and defect energies are provided in the Supplementary Material \cite{supplementary}.

\begin{figure}[bp]
\includegraphics[width=1 \linewidth]{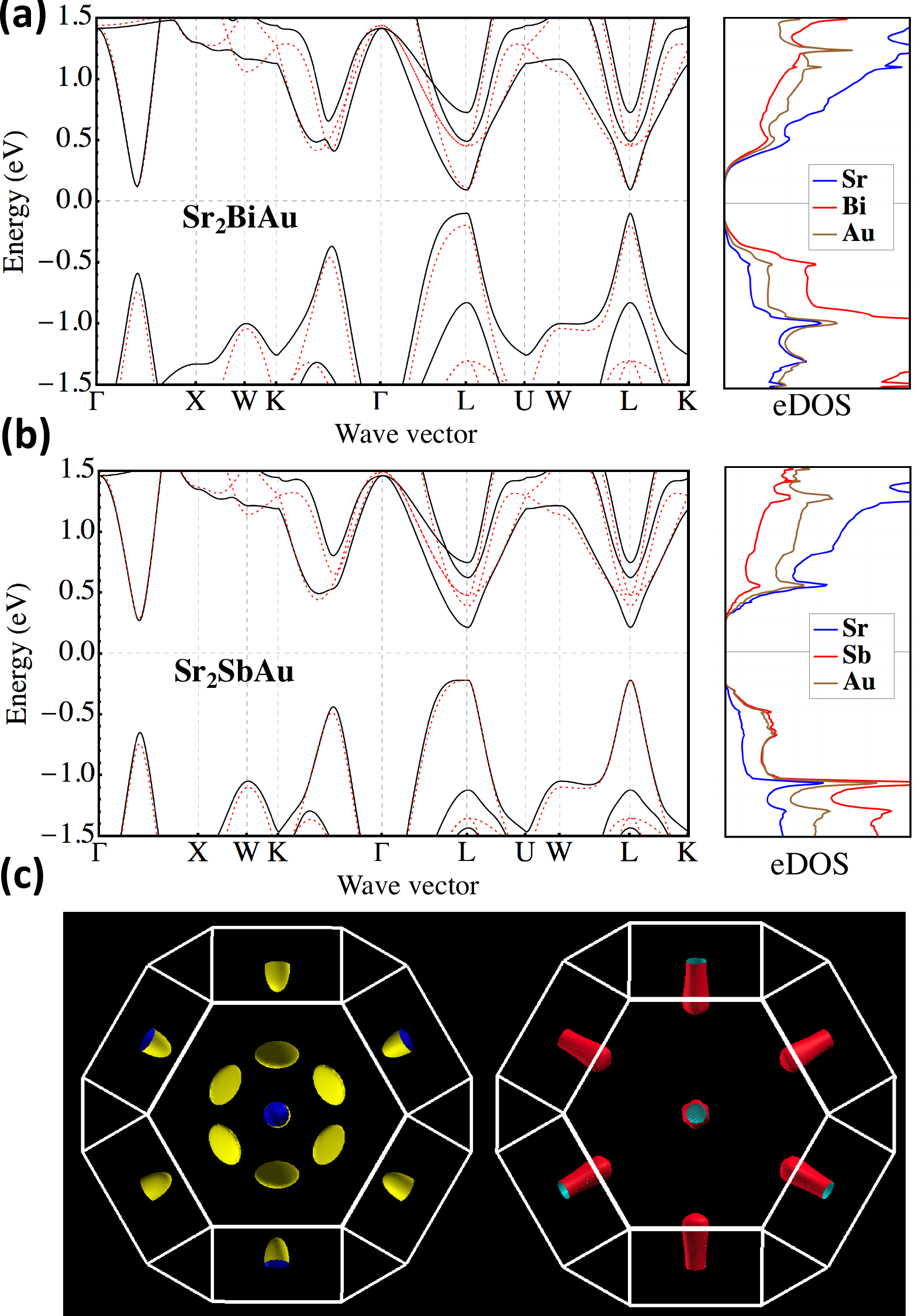}
\caption{(Color online) \textbf{a)} Electronic band structures of Sr$_{2}$BiAu with (black, solid) and without (red, dotted) SOC, aligned at the CBM. The atom-decomposed density of states with SOC is shown on the right. \textbf{b)} Same for Sr$_{2}$SbAu. \textbf{c)} Isoenergy surfaces of Sr$_{2}$SbAu with SOC, at 0.1 eV above the CBM (left) and below the VBM (right). The levels correspond to electron doping concentration of $n_{e}=1.44\times10^{20}$ cm$^{-3}$ and hole doping concentration of $n_{h}=1.40\times10^{20}$ cm$^{-3}$, respectively.} 
\label{fig:electronicstructure}
\end{figure}

\begin{figure*}[tp]
\includegraphics[width=1 \linewidth]{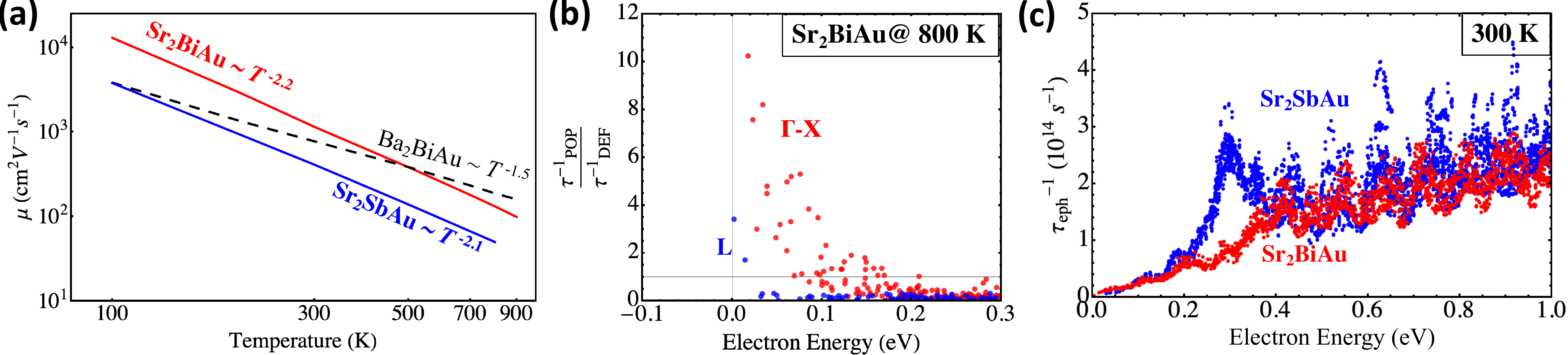}
\caption{(Color online) \textbf{a)} Comparison of electron mobilities of the Heusler compounds at low and high temperatures. \textbf{b)} The relative strength of polar-optical scattering (POP) and lattice deformation scattering (DEF) of the $\Gamma-X$-pocket (red), which is clearly dominated by POP, and the $L$-pocket (blue) which is heavily affected by DEF. \textbf{c)} Relative e-ph scattering rates of Sr$_{2}$BiAu and  Sr$_{2}$SbAu at 300 K.} 
\label{fig:eph}
\end{figure*}

\section{Electronic Properties}

\subsection{The Electronic Structure}

The band structures of Sr$_{2}$BiAu and  Sr$_{2}$SbAu are shown in Fig. \ref{fig:electronicstructure}a--b. Sr$_{2}$BiAu and Sr$_{2}$SbAu feature additional dispersive conduction band pockets at the fourfold degenerate $L$-point while retaining the sixfold degenerate pocket along $\Gamma-X$. SOC does not affect the conduction band pocket along  $\Gamma-X$, just as in the case of Ba$_{2}$BiAu. The corresponding energy surfaces as seen in Fig. \ref{fig:electronicstructure}c reveal all ten pockets. Whereas the $L$-pocket of Sr$_{2}$BiAu is nearly energy-aligned with the $\Gamma-X$-pocket, the $L$-pocket of Sr$_{2}$SbAu is lower than the $\Gamma-X$-pocket by 0.06 eV. The effective masses of the very dispersive $\Gamma-X$-pocket are essentially identical for all three compounds at approximately $m_{\parallel}=0.067$ and $m_{\perp}=0.48$. The $L$-pockets are somewhat less dispersive, and their principal effective masses are approximately $m_{\parallel}=0.19$ and $m_{\perp}=0.45$ for both of the Sr-compounds. 

A popular index for correlating a band structure to the PF it generates is the so-called Fermi surface complexity factor, calculated as \cite{complexityfactor}
\begin{equation}\label{eq:complexity}
C=N_{\text{pocket}}\left(\frac{2}{3}\left(\frac{m_{\perp}}{m_{\parallel}}\right)^{\frac{1}{3}}+\frac{1}{3}\left(\frac{m_{\perp}}{m_{\parallel}}\right)^{-\frac{2}{3}}\right)^{3/2},
\end{equation}
The band characters as described above yield complexity factors of 9.5 for Ba$_{2}$BiAu, which only has a $\Gamma-X$-pocket, and 14 for the two Sr-compounds, though the value is ambiguous for Sr$_{2}$SbAu where the pockets are misaligned. 

The band gaps as calculated by PBE+SOC are 0.19 eV for Sr$_{2}$BiAu and 0.5 eV for Sr$_{2}$SbAu, which are severe underestimations. The mBJ functional with SOC yields 0.53 eV Sr$_{2}$BiAu and 0.85 eV for Sr$_{2}$SbAu, where  HSE06 with SOC yields 0.53 eV and 0.81 eV, respectively.

\subsection{Scattering \& Mobility}

\begin{figure*}[tp]
\includegraphics[width=1 \linewidth]{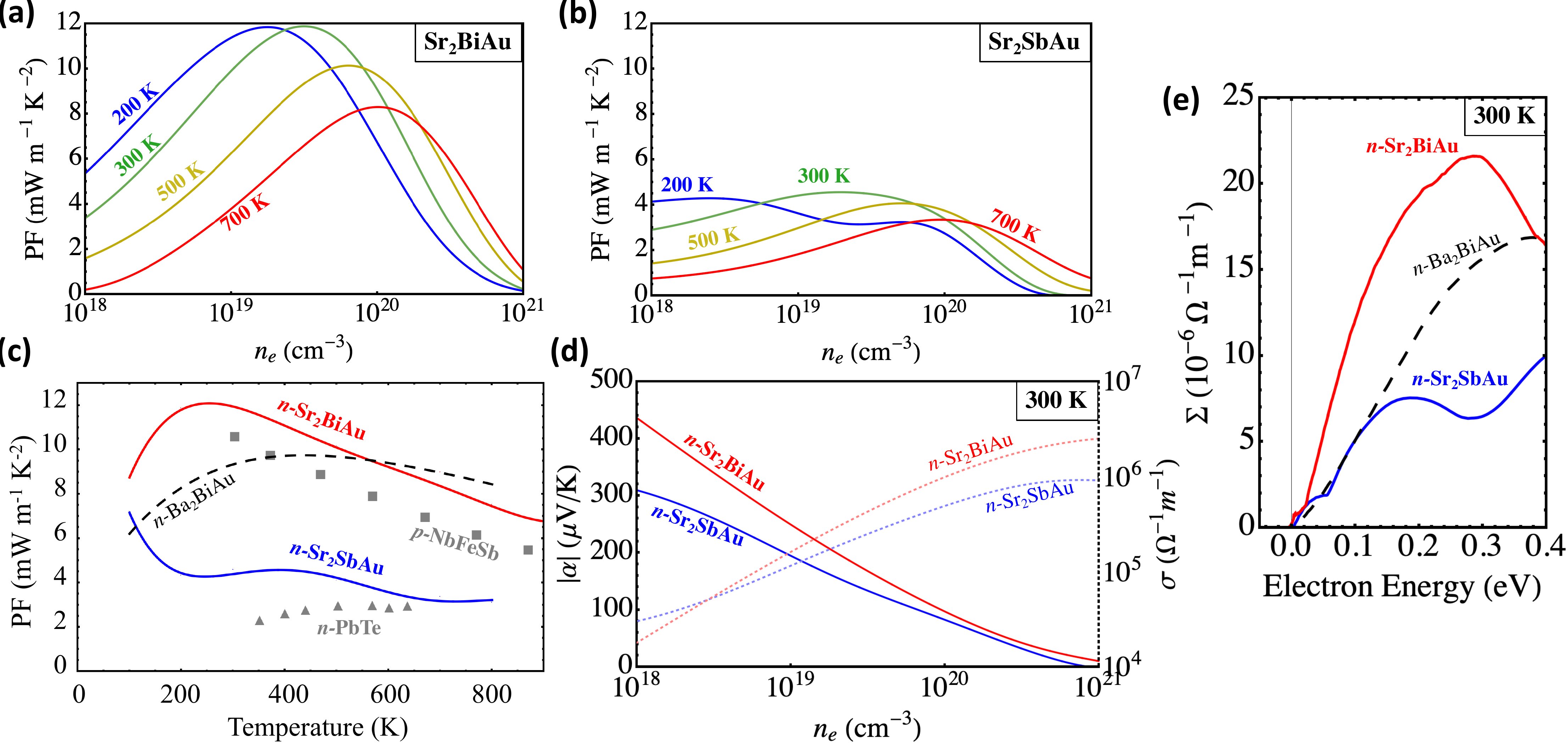}
\caption{(Color online) The electron-phonon-scattering-limited $n$-type power factor of \textbf{a)} Sr$_{2}$BiAu and \textbf{b)} Sr$_{2}$SbAu against electron doping concentration. \textbf{c)} Maximum $n$-type power factors of Sr$_{2}$BiAu and Sr$_{2}$BiAu at each temperature juxtaposed with power factors of high-performing thermoelectrics: $p$-type NbFeSb \cite{nbfesbpnas}, $n$-type PbTe \cite{pbtenanostructuring}, and theoretical $n$-type Ba$_{2}$BiAu \cite{ba2biau}). \textbf{d)} Comparison of the $n$-type Seebeck coefficient (solid) and conductivity (dotted) of Sr$_{2}$BiAu (red) and Sr$_{2}$SbAu (blue) at 300 K. \textbf{e)} The spectral conductivities of three full-Heusler compounds at 300 K, where they are all zero-aligned to their respective CBMs.}
\label{fig:thermoelectric}
\end{figure*}

Electron mobilities of the three Heusler compounds, calculated using band-and-\textbf{k}-dependent electron-phonon lifetimes, are juxtaposed in Fig. \ref{fig:eph}a. Electron mobilities of the two Sr compounds decay at a faster rate with temperature than that of Ba$_{2}$BiAu. This is due to the extra pocket $L$-pocket that the Sr compounds have. The $L$-pocket is overall heavier than the $\Gamma-X$-pocket and attains a larger electronic density of states (eDOS). Whereas the $\Gamma-X$-pocket is dispersive enough that the phase space for lattice deformation scattering is small, allowing polar-optical scattering to dominate, the heavier $L$-pocket is much more affected by deformation scattering [see Fig. \ref{fig:eph}c]. It is well known that lattice deformation results in quicker temperature-decay of mobility than polar-optical interactions. Therefore, presence of the $L$-states result in faster decay of mobility with temperature than if only the $\Gamma-X$-pocket were present, which in turn results in quicker decay of mobilities of the Sr compounds than that of  Ba$_{2}$BiAu. 

Though the mobilities of the Sr compounds exhibit similar trends, they notably differ in magnitude. This is partially due to the fact that the heavier $L$-pocket is the true band minimum in Sr$_{2}$SbAu, whereas in Sr$_{2}$BiAu it is nearly perfectly aligned with the $\Gamma-X$-pocket. Sr$_{2}$SbAu also generally experiences somewhat heavier scattering especially around 0.3 eV above the CBM, as seen in Fig. \ref{fig:eph}c. The differences in the temperature-dependence of mobility arising from accidentally degenerate pockets that are disparate in character translate to the behavior of their thermoelectric properties of the compounds, as will be seen.

We find that there is little to no intervalley scattering between the two pockets. When we artificially remove the $L$-pockets such that their participation to scattering of other states is forbidden, we detect virtually no change in the scattering rates of the remaining $\Gamma-X$-pocket.

\section{Thermoelectric Properties}

\subsection{The Power Factor}

Sr$_{2}$BiAu is capable of attaining very high $n$-type PFs across all temperatures, as shown in Fig. \ref{fig:thermoelectric}a, topping out at 12 mW m$^{-1}$K$^{-2}$ near room temperature. Sr$_{2}$SbAu simply performs worse, as made clear by Fig. \ref{fig:thermoelectric}b. It is a well-known engineering strategy to boost the PF to engineer bands of multiple pockets for energy convergence \cite{pbtebandconvergence,bandconvergencereview}. In essence Sr$_{2}$BiAu is a natural realization of this concept. Its theoretical PF limited by e-ph scattering hovers above the measured PF of the $p$-type NbFeSb \cite{nbfesbpnas} across all temperature domains [see Fig. \ref{fig:thermoelectric}c]. In contrast, Sr$_{2}$SbAu falls short of such natural band convergence as the heavier $L$-pocket is lower than the $\Gamma-X$-pocket by 0.06 eV. This, together with somewhat heavier scattering in Sr$_{2}$SbAu, results in both lower $\sigma$ and $\alpha$ for lower PF in comparison to Sr$_{2}$BiAu [see Fig. \ref{fig:thermoelectric}d.]

Comparison of Sr$_{2}$BiAu and Ba${_2}$BiAu provides insights to the effects of the additional, heavier $L$-pocket and the aforementioned temperature-dependent mobility profiles. At low temperatures, the PFs behave as expected from the complexity factors, and Sr$_{2}$BiAu easily performs better than Ba$_{2}$BiAu. However as higher temperatures excite deeper $L$ states whose lifetimes decay more quickly with energy (as reflected by mobilties), Ba$_{2}$BiAu begins to outperform Sr$_{2}$BiAu. Such a crossover indicates that the additional presence of a heavier pocket, under larger influence of lattice deformation scattering, benefits the PF only below certain threshold temperature past which the deep heavy states with short lifetimes are critically excited, negating the benefit of higher carrier population per Fermi level. The threshold temperature is, in turn, dependent upon the extent to which the second pocket is heavier than the first.

The essence of the overall relationship between the three band structures and their thermoelectric performance is represented by their energy-dependent spectral conductivities, $\Sigma(E)=D(E)\tau(E)v^{2}(E)$, plotted in Fig. \ref{fig:thermoelectric}e. In comparison to Ba$_{2}$BiAu, Sr$_{2}$BiAu attains noticeably steeper slope and higher values of $\Sigma(E)$ at its CBM due to simultaneous excitation of the $\Gamma-X$-and-$L$-pockets. High $\Sigma(E)$ with steep onset is a signature for high PF. Meanwhile, the profile for Sr$_{2}$SbAu is kinked: the main incline corresponding to the $\Gamma-X$-pocket occurs 0.06 eV into the CBM followed by a much more gradual onset corresponding to the heavier $L$-pocket, and it dips early around 0.3 eV due to the heavy scattering there. These features are less effective for thermoelectricity.

The main lesson of the above discussions is that a multitude of inherently distinct, symmetry-inequivalent pockets of accidental degeneracy is not necessarily beneficial for thermoelectrics. It better benefits thermoelectric performance if the pockets share similar dispersion and scattering behaviors. If sufficiently different in character, then one may be better off without the heavier pocket as the disparity in pocket lifetimes and mobilities would overpower increased carrier population. Only in the perfectly symmetry-identical cases do more pockets universally lead to higher performance. More generally, this demonstrates that indicators such as the complexity factor are valid in so far as all pockets share similar if not symmetry-identical profiles in not only the band shapes but also scattering behaviors. The complexity factor becomes an increasingly poorer measure of thermoelectric performance as the band pockets are accidentally degenerate and become more disparate in character.

\subsection{Thermal Properties and the Figure of Merit}

\begin{figure*}[tp]
\includegraphics[width=1 \linewidth]{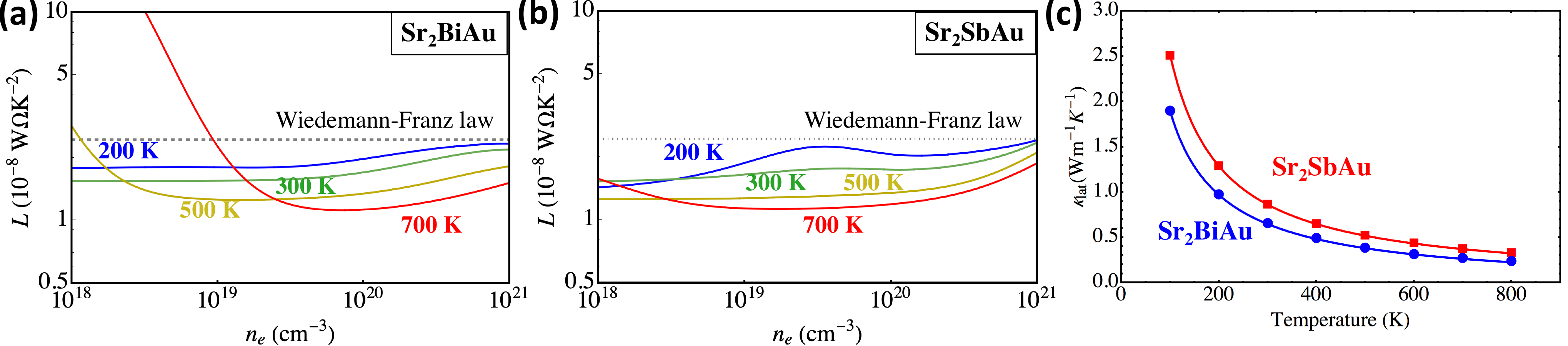}
\caption{(Color online) \textbf{a)} The Lorenz number of Sr$_{2}$BiAu against $n$-doping concentration. \textbf{b)} The Lorenz number of Sr$_{2}$SbAu against the Fermi level. The dotted horizontal lines mark the Wiedemann-Franz value. \textbf{c)} Lattice thermal conductivities of Sr$_{2}$BiAu and Sr$_{2}$SbAu.}
\label{fig:thermal}
\end{figure*}

\begin{figure*}[tp]
\includegraphics[width=0.8 \linewidth]{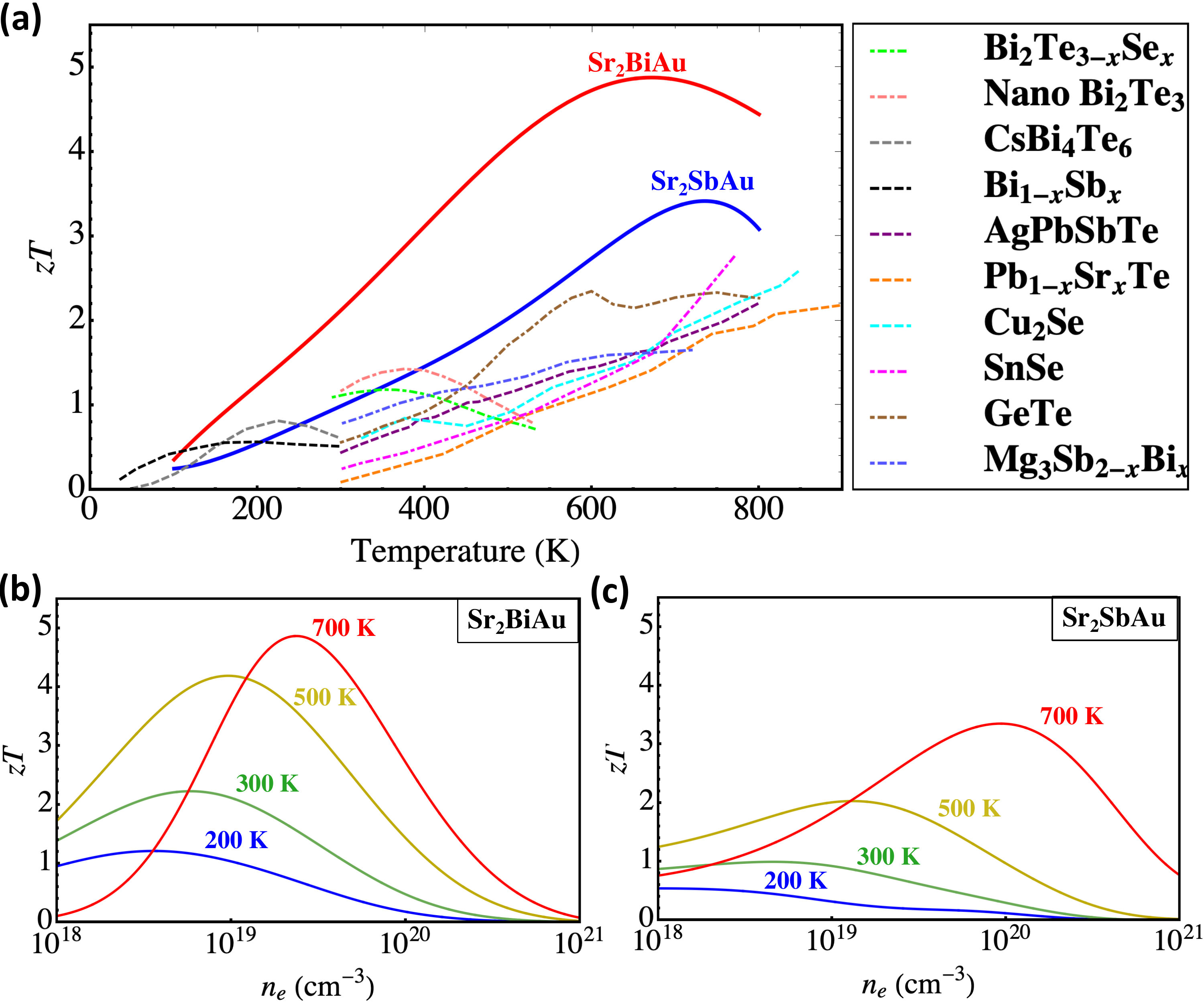}
\caption{(Color online) Theoretical $n$-type $zT$ of \textbf{a)} Sr$_{2}$BiAu and \textbf{b)} Sr$_{2}$SbAu against electron doping concentration. \textbf{c)} Theoretical \textit{maximum} $zT$ of $n$-type Sr$_{2}$BiAu and Sr$_{2}$SbAu at each temperature in comparison to state-of-the-art thermoelectrics \cite{last,bisbtenano,bitese,bisbtedislocation,cu2seuher,pbsrte,snsescience2,bisbsmithwolfe,gesbinte,getejoule,lowtmg3sb2,mg3sb2revelation}}
\label{fig:zt}
\end{figure*}

The Lorenz numbers ($L$) of the Sr-compounds are consistently below the free-electron Wiedemann-Franz value ($L_{\text{WF}}=2.44\times10^{-8}$ W$\Omega$K$^{-2}$). $L<L_{\text{WF}}$ by itself is expected as it reflects transport dominated by phonon scattering, whether due to polar-optical or lattice deformation \cite{thermoelectrics}. Yet upon a closer examination, a couple of anomalies are spotted. $L$ in these two compounds decrease with temperature whereas the free electron picture predicts increasing $L$ (towards $L_{\text{WF}}$) with temperature \cite{thermoelectrics}. Moreover, $L$ reaches as low as $10^{-8}$ W$\Omega$K$^{-2}$ at high temperatures for the two compounds. We attribute these results to the rapid rate at which scattering rates increase as the dominant process shifts from polar-optical near the band edge to lattice deformation in the deep electronic states. Referring back to Fig. \ref{fig:thermoelectric}e, about $0.3\sim0.35$ eV above the CBMs, $\Sigma(E)$ drops in magnitude after a peak, which is associated with the spike in the scattering rate and eDOS there. High-energy electrons occupying deep states contribute much more to $\kappa_{e}$ whereas low-energy electrons contribute more to $\sigma$. Therefore, comparatively faster decay of lifetimes at high energies, which are increasingly excited at higher temperatures, leads to lower $L$ than if just one scattering mechanism prevailed throughout.

The small magnitudes of $L$ and therefore $\kappa_{e}$ relative to $\sigma$ are particularly important because $\kappa_{e}>\kappa_{\text{lat}}$ in these compounds. The ultralow $\kappa_{\text{lat}}$ has been predicted by a previous study \cite{ultralowheusler}, whose results we reproduce here in Fig. \ref{fig:thermal}c using the same computational methods combining compressive sensing lattice dynamics \cite{csld} and iterative Boltzmann transport \cite{shengbte}. Phonon dispersion and density of states can be found in the Supplemental Material \cite{supplementary}.

The combination of high PF, and low $\kappa_{\text{lat}}$ and $L$ result in very high intrinsic $zT$ from cryogenic to high temperatures, especially for Sr$_{2}$BiAu. Notable in particular is the high performance $zT=0.4-2.2$ in the $100-300$ K range, which if realized would fill a niche at low temperatures. Comparisons made in Fig. \ref{fig:zt}a reveal that the theoretical performance of Sr$_{2}$BiAu, optimal doping assumed, is record-high at nearly all temperatures for bulk materials. Though not quite as glamorous, Sr$_{2}$SbAu is still poised to offer higher $zT$ than most of state-of-the-art thermoelectric compounds. If the offset $\Gamma-X$-and-$L$-pockets could be made to align in energy via doping or temperature effect, it would attain even higher PF and $zT$. Overall, because $\kappa_{e}$ is the dominant thermal conductivity and $L$ is rather constant, $zT$ tends to peak at lower $E_{\text{F}}$ ($n_{e}$) than the PF where higher Seebeck coefficient develops. This in turn reduces $n$-doping requirement to achieve optimum performance.

\section{Stability, Defects, and Dopability}

For the Heusler compounds hereby studied to experimentally realize their thermoelectric potentials, they ought to have a large region of stability and be $n$-dopable -- desirably to their ideal carrier concentrations of $n_{e}\approx10^{19}$ cm$^{-3}$. With the rapid emergence of computationally discovered hypothetical materials, it is important that analyses of realizability accompany performance predictions for a better guidance to experimentalists \cite{fantasymaterials}. Accordingly we analyze the stability and intrinsic defect energetics of the four full-Heusler compounds (Ba$_{2}$SbAu in addition to the three compounds studied above). 

\begin{figure}[bp]
\includegraphics[width=0.8 \linewidth]{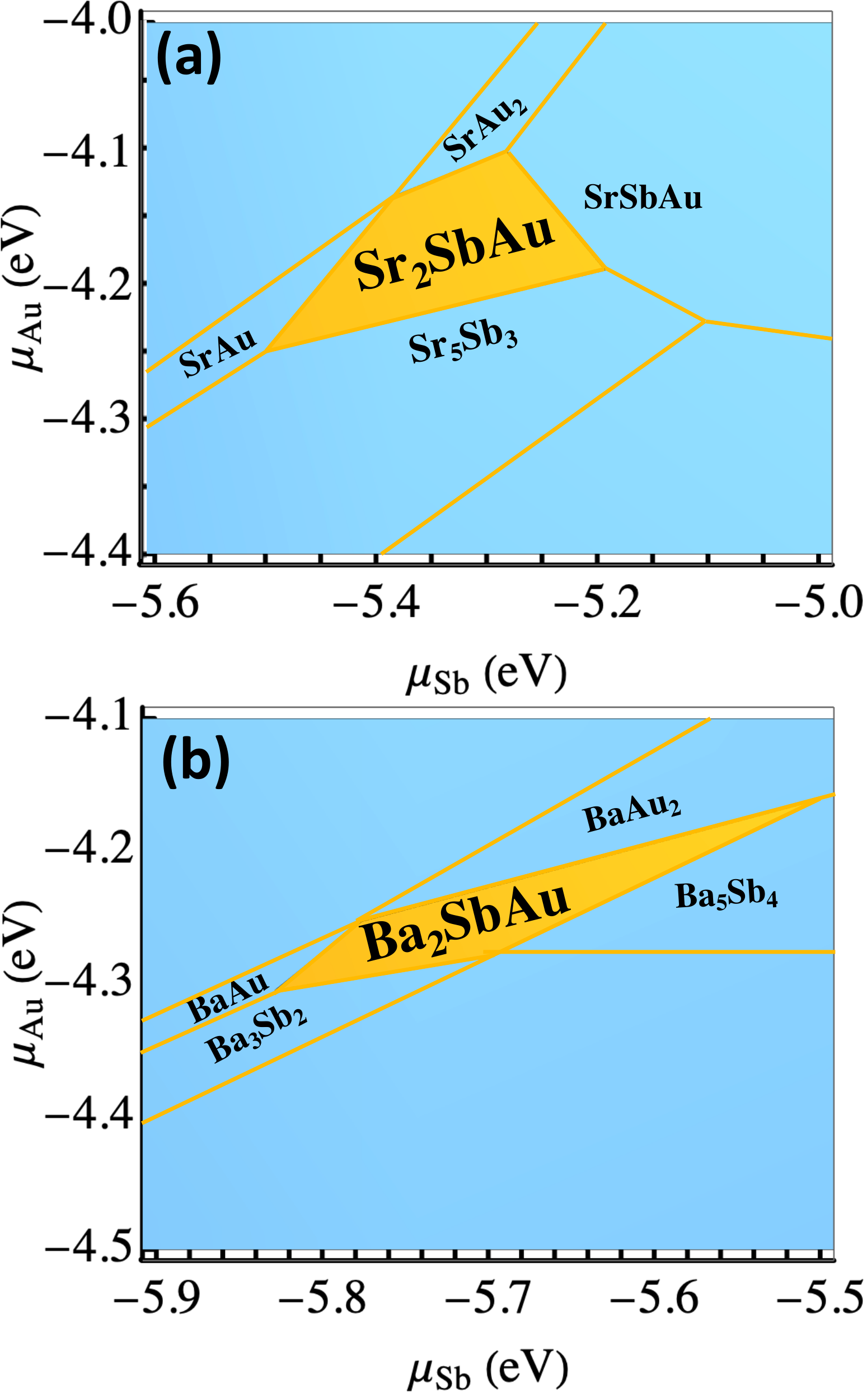}
\caption{(Color online) The region of phase stability (shaded orange) of \textbf{a)} Sr$_{2}$SbAu and \textbf{b)} Ba$_{2}$SbAu in the chemical potential space of Sb (the horizontal axes), and Au (the vertical axes). Secondary competing phases to be found in the vicinity are labeled.} 
\label{fig:chempot}
\end{figure}

\subsection{Phase Stability}

We find that the two Bi-compounds, Sr$_{2}$BiAu and Ba$_{2}$BiAu, are barely thermodynamically stable, as in that they have extremely limited chemical potential spaces of phase stability - essentially single points. Their points of stability are $\mu_{\text{Sr}}\approx-2.66$ eV, $\mu_{\text{Bi}}\approx-5.26$ eV, and $\mu_{\text{Au}}\approx-4.13$ for Sr$_{2}$BiAu, and $\mu_{\text{Ba}}\approx-2.78$ eV, $\mu_{\text{Bi}}\approx-5.71$ eV, and $\mu_{\text{Au}}\approx-4.25$ for Ba$_{2}$BiAu. This is a rather unfavorable sign for their synthesis. On the other hand, the two Sb-compounds, Sr$_{2}$SbAu and Ba$_{2}$SbAu, are found to have sizable regions of phase stability as shown in Fig. \ref{fig:chempot}, and therefore are better poised to be synthesized experimentally. It is also worth mentioning that a $P6_{3}mmc$ phase of 1:1:1 stoichiometry have been experimentally observed \cite{rmxcompounds}, which may serve as a starting point for tuning the chemical potentials.

\begin{figure}[bp]
\includegraphics[width=0.85 \linewidth]{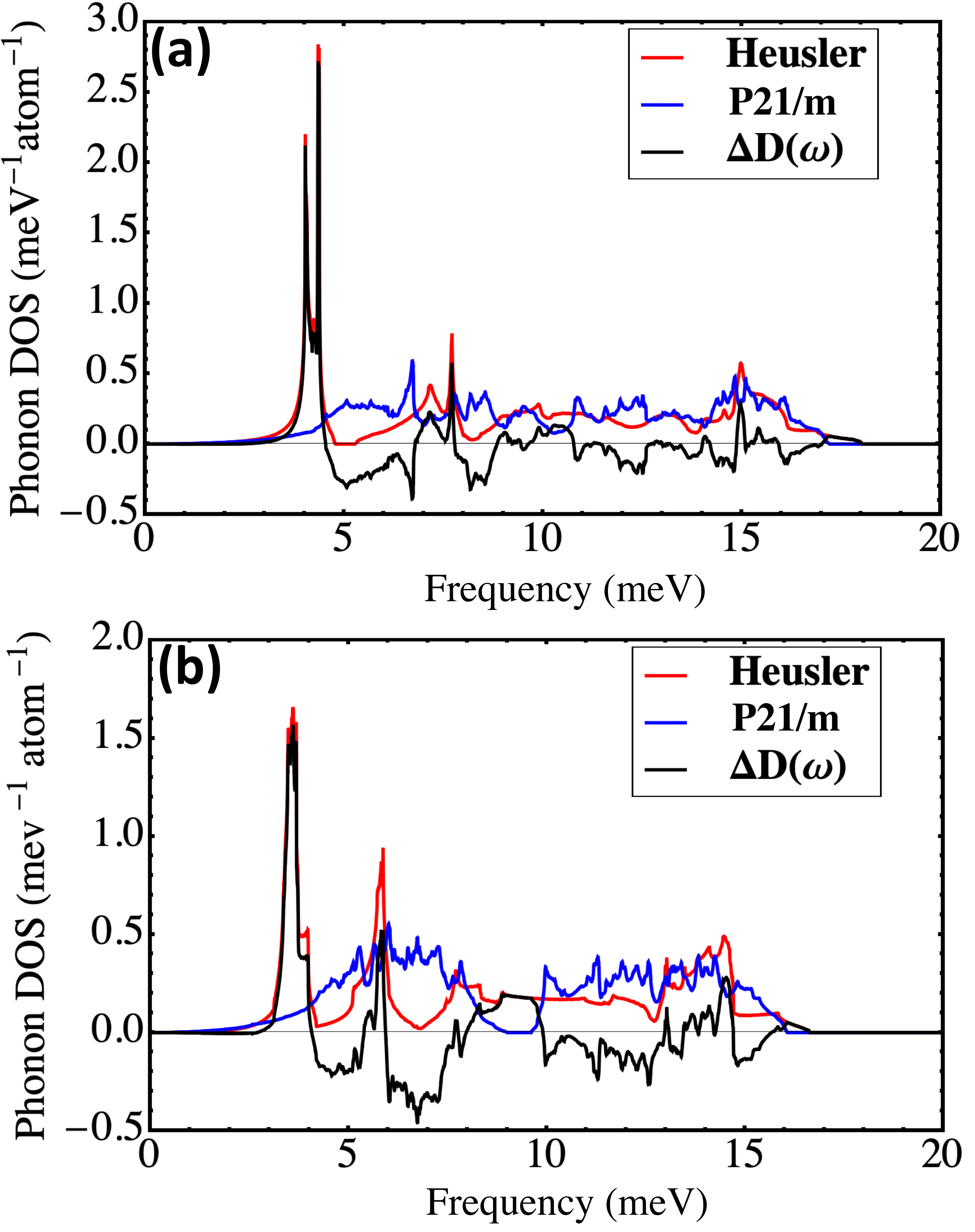}
\caption{(Color online) The phonon DOS of the Heusler and $P21/m$ phases and their differences for \textbf{a)} Sr$_{2}$SbAu and \textbf{b)} Sr$_{2}$BiAu.}
\label{fig:deltados}
\end{figure}

The two Sr-compounds, Sr$_{2}$BiAu and Sr$_{2}$SbAu, each have a stoichiometrically identical polymorphs of the $P21/m$ phase, as brought up by a previous work \cite{ultralowheusler}. As per our PBE+SOC total energy calculation, these phases are slightly lower in energy than the Heusler counterparts, by approximately 4 meV and 10 meV per atom respectively. However, these small differences are beyond the numerical resolution of DFT, and either polymorph would have a chance of forming. At finite temperatures, the relative stability would thus likely boil down to the entropic contributions to the free energy. We find that the Heusler phases have much higher vibrational entropies ($S_{v}$) than the $P21/m$ phases, by +0.42 $k_{\text{B}}$ per atom for Sr$_{2}$BiAu and +0.36 $k_{\text{B}}$ per atom for Sr$_{2}$SbAu. 
Fig. \ref{fig:deltados} shows that the prominently soft acoustic modes of the Heusler phases, represented by the much higher net phonon DOS ($\Delta D(\omega)$) in the low-frequency region, drive their higher $S_{v}$. At 300 K, the net gain in $\Delta S_{v}$ of the Heusler phases lowers their free energy of formation by 12 meV and 9 meV per atom, respectively, relative to the $P21/m$ phases. As a result at 300 K, Heusler Sr$_{2}$BiAu becomes stable by 8 meV per atom and Heusler Sr$_{2}$SbAu also turns essentially stable.

\subsection{Defects}

\begin{figure*}[tp]
\includegraphics[width=1 \linewidth]{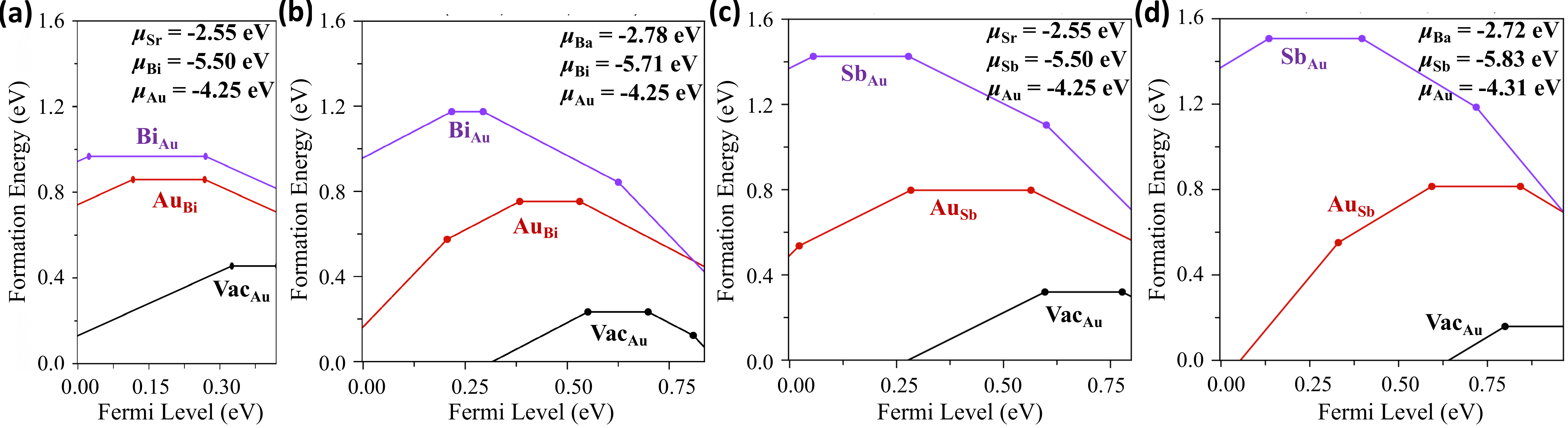}
\caption{(Color online) The formation energies of three lowest-energy defects Vac$_{\text{Au}}$, Au$_{\text{Bi/Sb}}$, and Bi/Sb$_{\text{Au}}$, in \textbf{a)} Sr$_{2}$BiAu, \textbf{b)} Ba$_{2}$BiAu, \textbf{c)} Sr$_{2}$SbAu, and \textbf{d)} Ba$_{2}$SbAu. The slopes of the lines indicate charge ($q=-2\sim+2$). The plots go from the VBM (0 eV) to the CBM as determined by HSE+SOC. The chemical potentials for the plots shown correspond to the Au-poor conditions, where $\mu_{\text{Au}}$ is the lowest, at which Vac$_{\text{Au}}$ formation energy is minimized.}
\label{fig:defects}
\end{figure*}

In all four compounds, the lowest-energy defect is Au vacancy (Vac$_{\text{Au}}$). This likely reflects that, at their sites, Au atoms are very weakly bonded in both compounds. It has been discussed in Ref. \onlinecite{ultralowheusler} that due to its filled $5d$ and $6s$ shells, loosely bonded Au atoms act as rattlers with strongly anharmonic vibrations and are key contributors to phonon scattering and very low $\kappa_{\text{lat}}$ in the two compounds. The formation energies are shown in Fig. \ref{fig:defects} along with those of the next-highest-energy defects, namely the Au$_{\text{Bi}}$ and Bi$_{\text{Au}}$ antisites. The small magnitudes of the correction term $E_{\text{cor}}$ (less than 0.1 eV for most charged defects for all types) reflect that high $\epsilon$ of the compounds inhibit strong electrostatic interactions between charged defects and their periodic images, validating the overall size-sufficiency of the supercells used.

We obtain the defect-limited Fermi levels by enforcing charge neutrality of all defects. Zero-referenced to the VBMs, the values are 0.62 eV, 0.35 eV, 0.69 eV, and 0.91 eV respectively for Ba$_{2}$BiAu, Sr$_{2}$BiAu, Sr$_{2}$SbAu, and Ba$_{2}$SbAu. As these are in all four cases close to the conduction band edges (0.82 eV, 0.42 eV, 0.80 eV, 0.95 eV, respectively), Vac$_{\text{Au}}$ intrinsically drives the Heusler systems to be somewhat $n$-type. However, these defect-pinned Fermi levels correspond to excess electron concentrations of only $10^{15}\sim10^{18}$ cm$^{-3}$ at 300 K [see Fig. S5 in Supplemental Material \cite{supplementary}]. This is not quite sufficient for reaching the optimal doping concentrations of $10^{18}\sim10^{20}$ cm$^{-3}$. Further excess electrons then must be introduced externally. We find that Hg$_{\text{Au}}$, a potent external $n$-dopant, is stable by more than 100 meV per atom with respect to the convex hull of all other known stable phases formed by the host atoms and Hg. Considering also that Hg is just as heavy as Au, its introduction is expected to preserve the Au-dominated soft acoustic phonons, thus maintaining high positive $S_{v}$ and low $\kappa_{\text{lat}}$. For these reasons, toxicity notwithstanding, Hg$_{\text{Au}}$ would be a strategic doping mechanism for optimal triggering of $n$-type performance.

\section{Conclusion}

In summary, the full-Heuslers in this study - especially Sr$_{2}$BiAu - constitute a theoretical validation that very high \textit{intrinsic} thermoelectric performance across a broad range of temperatures from $100\sim800$ K is physically possible in real bulk compounds. The predicted $zT$ values are notably high at cryogenic-to-room temperatures -- the domain that generally lacks efficient thermoelectric materials. Dispersive conduction band pockets at two off-symmetry points (for a total of ten pockets in the Brillouin zone) generate very high power factors across all temperatures. Notably, due to the disparity in the characters of the two pockets in terms of dispersion and dominant scattering mechanisms, the power factor decays more quickly with temperature. Whether accidentally degenerate pockets benefit thermoelectric performance is therefore conditional upon the similarity of the pockets and temperature of operation. If the pockets are too dissimilar, then performance would benefit if the lighter pocket were alone without the heavier pocket.

Fortunately, intrinsic defect energetics is favorable for $n$-doping. Au vacancy has the lowest formation energy of all defects:  $0.2\sim0.4$ eV for charge-neutral vacancy. All things considered, the two Sb-compounds, Sr$_{2}$SbAu and Ba$_{2}$SbAu, have the best chance at realization since they have the largest chemical potential regions of phase stability. With the help of external $n$-dopants such as Hg$_{\text{Au}}$, near-optimal triggering of their thermoelectricity may well be within experimental reach. The Bi-compounds, Sr$_{2}$BiAu and Ba$_{2}$BiAu, appear to be fundamentally hampered by their constricted stability conditions. Yet, at high temperatures where the effects of vibrational entropy increasingly favors formation of the Heusler phases over others, their realization may stand a better chance.

\begin{acknowledgements}
The study was conceived and designed on National Science Foundation (NSF) Grant DMR-1611507, which funded V.O. The study was further developed and completed on the grant from U.S. Department of Energy, Office of Basic Energy Sciences, Early Career Research Program, which funded J. P., A. G., and A. J. This research used computational resources of 1) the National Energy Research Scientific Computing Center (NERSC), a DOE Office of Science User Facility supported by the Office of Science of the U.S. Department of Energy under Contract No. DE-AC02-05CH11231, and 2) Texas Advanced Computing Center (TACC) at the University of Texas at Austin through the Extreme Science and Engineering Discovery Environment (XSEDE), which is supported by NSF grant number ACI-1548562.
\end{acknowledgements}

\bibliography{references}

\begin{thebibliography}{90}%
\makeatletter
\providecommand \@ifxundefined [1]{%
 \@ifx{#1\undefined}
}%
\providecommand \@ifnum [1]{%
 \ifnum #1\expandafter \@firstoftwo
 \else \expandafter \@secondoftwo
 \fi
}%
\providecommand \@ifx [1]{%
 \ifx #1\expandafter \@firstoftwo
 \else \expandafter \@secondoftwo
 \fi
}%
\providecommand \natexlab [1]{#1}%
\providecommand \enquote  [1]{``#1''}%
\providecommand \bibnamefont  [1]{#1}%
\providecommand \bibfnamefont [1]{#1}%
\providecommand \citenamefont [1]{#1}%
\providecommand \href@noop [0]{\@secondoftwo}%
\providecommand \href [0]{\begingroup \@sanitize@url \@href}%
\providecommand \@href[1]{\@@startlink{#1}\@@href}%
\providecommand \@@href[1]{\endgroup#1\@@endlink}%
\providecommand \@sanitize@url [0]{\catcode `\\12\catcode `\$12\catcode
  `\&12\catcode `\#12\catcode `\^12\catcode `\_12\catcode `\%12\relax}%
\providecommand \@@startlink[1]{}%
\providecommand \@@endlink[0]{}%
\providecommand \url  [0]{\begingroup\@sanitize@url \@url }%
\providecommand \@url [1]{\endgroup\@href {#1}{\urlprefix }}%
\providecommand \urlprefix  [0]{URL }%
\providecommand \Eprint [0]{\href }%
\providecommand \doibase [0]{http://dx.doi.org/}%
\providecommand \selectlanguage [0]{\@gobble}%
\providecommand \bibinfo  [0]{\@secondoftwo}%
\providecommand \bibfield  [0]{\@secondoftwo}%
\providecommand \translation [1]{[#1]}%
\providecommand \BibitemOpen [0]{}%
\providecommand \bibitemStop [0]{}%
\providecommand \bibitemNoStop [0]{.\EOS\space}%
\providecommand \EOS [0]{\spacefactor3000\relax}%
\providecommand \BibitemShut  [1]{\csname bibitem#1\endcsname}%
\let\auto@bib@innerbib\@empty
\bibitem [{\citenamefont {Olvera}\ \emph {et~al.}(2017)\citenamefont {Olvera},
  \citenamefont {Moroz}, \citenamefont {Sahoo}, \citenamefont {Ren},
  \citenamefont {Bailey}, \citenamefont {Page}, \citenamefont {Uher},\ and\
  \citenamefont {Poudeu}}]{cu2seuher}%
  \BibitemOpen
  \bibfield  {author} {\bibinfo {author} {\bibfnamefont {A.~A.}\ \bibnamefont
  {Olvera}}, \bibinfo {author} {\bibfnamefont {N.~A.}\ \bibnamefont {Moroz}},
  \bibinfo {author} {\bibfnamefont {P.}~\bibnamefont {Sahoo}}, \bibinfo
  {author} {\bibfnamefont {P.}~\bibnamefont {Ren}}, \bibinfo {author}
  {\bibfnamefont {T.~P.}\ \bibnamefont {Bailey}}, \bibinfo {author}
  {\bibfnamefont {A.~A.}\ \bibnamefont {Page}}, \bibinfo {author}
  {\bibfnamefont {C.}~\bibnamefont {Uher}}, \ and\ \bibinfo {author}
  {\bibfnamefont {P.~F.~P.}\ \bibnamefont {Poudeu}},\ }\bibfield  {title}
  {Partial indium solubility induces chemical stability and colossal
  thermoelectric figure of merit in Cu$_{2}$Se,\ }\href@noop {} {\bibfield
  {journal} {\bibinfo  {journal} {Energy Environ. Sci..}\ }\textbf {\bibinfo
  {volume} {10}},\ \bibinfo {pages} {1668--1676} (\bibinfo {year}
  {2017})}\BibitemShut {NoStop}%
\bibitem [{\citenamefont {Zhong}\ \emph {et~al.}(2014)\citenamefont {Zhong},
  \citenamefont {Zhang}, \citenamefont {Li}, \citenamefont {Chen},
  \citenamefont {Cui}, \citenamefont {Li}, \citenamefont {Xie}, \citenamefont
  {Hao},\ and\ \citenamefont {He}}]{alincuse}%
  \BibitemOpen
  \bibfield  {author} {\bibinfo {author} {\bibfnamefont {Bin}\ \bibnamefont
  {Zhong}}, \bibinfo {author} {\bibfnamefont {Yong}\ \bibnamefont {Zhang}},
  \bibinfo {author} {\bibfnamefont {Weiqian}\ \bibnamefont {Li}}, \bibinfo
  {author} {\bibfnamefont {Zhenrui}\ \bibnamefont {Chen}}, \bibinfo {author}
  {\bibfnamefont {Jingying}\ \bibnamefont {Cui}}, \bibinfo {author}
  {\bibfnamefont {Wei}\ \bibnamefont {Li}}, \bibinfo {author} {\bibfnamefont
  {Yuandong}\ \bibnamefont {Xie}}, \bibinfo {author} {\bibfnamefont {Qing}\
  \bibnamefont {Hao}}, \ and\ \bibinfo {author} {\bibfnamefont {Qinyu}\
  \bibnamefont {He}},\ }\bibfield  {title} {High superionic conduction arising
  from aligned large lamellae and large figure of merit in bulk
  Cu$_{1.94}$Al$_{0.02}$Se,\ }\href@noop {} {\bibfield  {journal} {\bibinfo
  {journal} {Appl. Phys. Lett.}\ }\textbf {\bibinfo {volume} {105}},\ \bibinfo
  {pages} {123902} (\bibinfo {year} {2014})}\BibitemShut {NoStop}%
\bibitem [{\citenamefont {Zhao}\ \emph {et~al.}(2016)\citenamefont {Zhao},
  \citenamefont {Tan}, \citenamefont {Hao}, \citenamefont {He}, \citenamefont
  {Pei}, \citenamefont {Chi}, \citenamefont {Wang}, \citenamefont {Gong},
  \citenamefont {Xu}, \citenamefont {Dravid}, \citenamefont {Uher},
  \citenamefont {Snyder}, \citenamefont {Wolverton},\ and\ \citenamefont
  {Kanatzidis}}]{snsescience1}%
  \BibitemOpen
  \bibfield  {author} {\bibinfo {author} {\bibfnamefont {L.}~\bibnamefont
  {Zhao}}, \bibinfo {author} {\bibfnamefont {G.}~\bibnamefont {Tan}}, \bibinfo
  {author} {\bibfnamefont {S.}~\bibnamefont {Hao}}, \bibinfo {author}
  {\bibfnamefont {J.}~\bibnamefont {He}}, \bibinfo {author} {\bibfnamefont
  {Y.}~\bibnamefont {Pei}}, \bibinfo {author} {\bibfnamefont {H.}~\bibnamefont
  {Chi}}, \bibinfo {author} {\bibfnamefont {H.}~\bibnamefont {Wang}}, \bibinfo
  {author} {\bibfnamefont {S.}~\bibnamefont {Gong}}, \bibinfo {author}
  {\bibfnamefont {H.}~\bibnamefont {Xu}}, \bibinfo {author} {\bibfnamefont
  {V.}~\bibnamefont {Dravid}}, \bibinfo {author} {\bibfnamefont
  {C.}~\bibnamefont {Uher}}, \bibinfo {author} {\bibfnamefont {G.~J.}\
  \bibnamefont {Snyder}}, \bibinfo {author} {\bibfnamefont {C.}~\bibnamefont
  {Wolverton}}, \ and\ \bibinfo {author} {\bibfnamefont {M.~G.}\ \bibnamefont
  {Kanatzidis}},\ }\bibfield  {title} {Ultrahigh power factor and
  thermoelectric performance in hole-doped single-crystal SnSe,\ }\href@noop {}
  {\bibfield  {journal} {\bibinfo  {journal} {Science}\ }\textbf {\bibinfo
  {volume} {351}},\ \bibinfo {pages} {141--144} (\bibinfo {year}
  {2016})}\BibitemShut {NoStop}%
\bibitem [{\citenamefont {Chang}\ \emph {et~al.}(2018)\citenamefont {Chang},
  \citenamefont {Wu}, \citenamefont {He}, \citenamefont {Pei}, \citenamefont
  {Wu}, \citenamefont {Wu}, \citenamefont {Yu}, \citenamefont {Zhu},
  \citenamefont {Wang}, \citenamefont {Chen}, \citenamefont {Huang},
  \citenamefont {Li}, \citenamefont {He},\ and\ \citenamefont
  {Zhao}}]{snsescience2}%
  \BibitemOpen
  \bibfield  {author} {\bibinfo {author} {\bibfnamefont {Cheng}\ \bibnamefont
  {Chang}}, \bibinfo {author} {\bibfnamefont {Minghui}\ \bibnamefont {Wu}},
  \bibinfo {author} {\bibfnamefont {Dongsheng}\ \bibnamefont {He}}, \bibinfo
  {author} {\bibfnamefont {Yanling}\ \bibnamefont {Pei}}, \bibinfo {author}
  {\bibfnamefont {Chao-Feng}\ \bibnamefont {Wu}}, \bibinfo {author}
  {\bibfnamefont {Xuefeng}\ \bibnamefont {Wu}}, \bibinfo {author}
  {\bibfnamefont {Hulei}\ \bibnamefont {Yu}}, \bibinfo {author} {\bibfnamefont
  {Fangyuan}\ \bibnamefont {Zhu}}, \bibinfo {author} {\bibfnamefont {Kedong}\
  \bibnamefont {Wang}}, \bibinfo {author} {\bibfnamefont {Yue}\ \bibnamefont
  {Chen}}, \bibinfo {author} {\bibfnamefont {Li}~\bibnamefont {Huang}},
  \bibinfo {author} {\bibfnamefont {Jing-Feng}\ \bibnamefont {Li}}, \bibinfo
  {author} {\bibfnamefont {Jiaqing}\ \bibnamefont {He}}, \ and\ \bibinfo
  {author} {\bibfnamefont {Li-Dong}\ \bibnamefont {Zhao}},\ }\bibfield  {title}
  {3D charge and 2D phonon transports leading to high out-of-plane ZT in
  $n$-type SnSe crystals,\ }\href@noop {} {\bibfield  {journal} {\bibinfo
  {journal} {Science}\ }\textbf {\bibinfo {volume} {360}},\ \bibinfo {pages}
  {778--783} (\bibinfo {year} {2018})}\BibitemShut {NoStop}%
\bibitem [{\citenamefont {Duong}\ \emph {et~al.}(2016)\citenamefont {Duong},
  \citenamefont {Nguyen}, \citenamefont {Duvjir}, \citenamefont {Duong},
  \citenamefont {Kwon}, \citenamefont {Song}, \citenamefont {Lee},
  \citenamefont {Lee}, \citenamefont {Park}, \citenamefont {Min}, \citenamefont
  {Lee}, \citenamefont {Kim},\ and\ \citenamefont {Cho}}]{snsenatcomm}%
  \BibitemOpen
  \bibfield  {author} {\bibinfo {author} {\bibfnamefont {Anh~Tuan}\
  \bibnamefont {Duong}}, \bibinfo {author} {\bibfnamefont {Van~Quang}\
  \bibnamefont {Nguyen}}, \bibinfo {author} {\bibfnamefont {Ganbat}\
  \bibnamefont {Duvjir}}, \bibinfo {author} {\bibfnamefont {Van~Thiet}\
  \bibnamefont {Duong}}, \bibinfo {author} {\bibfnamefont {Suyong}\
  \bibnamefont {Kwon}}, \bibinfo {author} {\bibfnamefont {Jae~Yong}\
  \bibnamefont {Song}}, \bibinfo {author} {\bibfnamefont {Jae~Ki}\ \bibnamefont
  {Lee}}, \bibinfo {author} {\bibfnamefont {Ji~Eun}\ \bibnamefont {Lee}},
  \bibinfo {author} {\bibfnamefont {SuDong}\ \bibnamefont {Park}}, \bibinfo
  {author} {\bibfnamefont {Taewon}\ \bibnamefont {Min}}, \bibinfo {author}
  {\bibfnamefont {Jaekwang}\ \bibnamefont {Lee}}, \bibinfo {author}
  {\bibfnamefont {Jungdae}\ \bibnamefont {Kim}}, \ and\ \bibinfo {author}
  {\bibfnamefont {Sunglae}\ \bibnamefont {Cho}},\ }\bibfield  {title}
  {Achieving $zT=2.2$ with Bi-doped $n$-type SnSe single crystals,\ }\href@noop
  {} {\bibfield  {journal} {\bibinfo  {journal} {Nat. Commun.}\ }\textbf
  {\bibinfo {volume} {7}} (\bibinfo {year} {2016})}\BibitemShut {NoStop}%
\bibitem [{\citenamefont {Hsu}\ \emph {et~al.}(2004)\citenamefont {Hsu},
  \citenamefont {Loo}, \citenamefont {Guo}, \citenamefont {Chen}, \citenamefont
  {Dyck}, \citenamefont {Uher}, \citenamefont {Hogan}, \citenamefont
  {Polychroniadis},\ and\ \citenamefont {Kanatzidis}}]{last}%
  \BibitemOpen
  \bibfield  {author} {\bibinfo {author} {\bibfnamefont {Kuei~Fang}\
  \bibnamefont {Hsu}}, \bibinfo {author} {\bibfnamefont {Sim}\ \bibnamefont
  {Loo}}, \bibinfo {author} {\bibfnamefont {Fu}~\bibnamefont {Guo}}, \bibinfo
  {author} {\bibfnamefont {Wei}\ \bibnamefont {Chen}}, \bibinfo {author}
  {\bibfnamefont {Jeffrey~S.}\ \bibnamefont {Dyck}}, \bibinfo {author}
  {\bibfnamefont {Ctirad}\ \bibnamefont {Uher}}, \bibinfo {author}
  {\bibfnamefont {Tim}\ \bibnamefont {Hogan}}, \bibinfo {author} {\bibfnamefont
  {E.~K.}\ \bibnamefont {Polychroniadis}}, \ and\ \bibinfo {author}
  {\bibfnamefont {Mercouri~G.}\ \bibnamefont {Kanatzidis}},\ }\bibfield
  {title} {Cubic AgPb$_{m}$SbTe$_{2+m}$: Bulk Thermoelectric Materials with
  High Figure of Merit,\ }\href@noop {} {\bibfield  {journal} {\bibinfo
  {journal} {Science}\ }\textbf {\bibinfo {volume} {303}},\ \bibinfo {pages}
  {818--821} (\bibinfo {year} {2004})}\BibitemShut {NoStop}%
\bibitem [{\citenamefont {Wu}\ \emph {et~al.}(2014)\citenamefont {Wu},
  \citenamefont {Zhao}, \citenamefont {Zheng}, \citenamefont {Wu},
  \citenamefont {Pei}, \citenamefont {Tong}, \citenamefont {Kanatzidis},\ and\
  \citenamefont {He}}]{pbtes}%
  \BibitemOpen
  \bibfield  {author} {\bibinfo {author} {\bibfnamefont {H.~J.}\ \bibnamefont
  {Wu}}, \bibinfo {author} {\bibfnamefont {L-D}\ \bibnamefont {Zhao}}, \bibinfo
  {author} {\bibfnamefont {F.S.}\ \bibnamefont {Zheng}}, \bibinfo {author}
  {\bibfnamefont {D.}~\bibnamefont {Wu}}, \bibinfo {author} {\bibfnamefont
  {Y.L.}\ \bibnamefont {Pei}}, \bibinfo {author} {\bibfnamefont
  {X.}~\bibnamefont {Tong}}, \bibinfo {author} {\bibfnamefont {M.G.}\
  \bibnamefont {Kanatzidis}}, \ and\ \bibinfo {author} {\bibfnamefont {J.~Q.}\
  \bibnamefont {He}},\ }\bibfield  {title} {Broad temperature plateau for
  thermoelectric figure of merit $ZT>2$ in phase-separated
  PbTe$_{0.7}$S$_{0.3}$,\ }\href@noop {} {\bibfield  {journal} {\bibinfo
  {journal} {Nat. Commun.}\ }\textbf {\bibinfo {volume} {5}},\ \bibinfo {pages}
  {1--7} (\bibinfo {year} {2014})}\BibitemShut {NoStop}%
\bibitem [{\citenamefont {Hong}\ \emph {et~al.}(2018)\citenamefont {Hong},
  \citenamefont {Chen}, \citenamefont {Yang}, \citenamefont {Zou},
  \citenamefont {Dargusch}, \citenamefont {Wang},\ and\ \citenamefont
  {Zou}}]{gesbinte}%
  \BibitemOpen
  \bibfield  {author} {\bibinfo {author} {\bibfnamefont {Min}\ \bibnamefont
  {Hong}}, \bibinfo {author} {\bibfnamefont {Zhi-Gang}\ \bibnamefont {Chen}},
  \bibinfo {author} {\bibfnamefont {Lei}\ \bibnamefont {Yang}}, \bibinfo
  {author} {\bibfnamefont {Yi-Chao}\ \bibnamefont {Zou}}, \bibinfo {author}
  {\bibfnamefont {Matthew~S.}\ \bibnamefont {Dargusch}}, \bibinfo {author}
  {\bibfnamefont {Hao}\ \bibnamefont {Wang}}, \ and\ \bibinfo {author}
  {\bibfnamefont {Jin}\ \bibnamefont {Zou}},\ }\bibfield  {title} {Realizing
  $zT$ of 2.3 in Ge$_{1-x-y}$Sb$_{x}$In$_{y}$Te via Reducing the
  Phase-Transition Temperature and Introducing Resonant Energy Doping,\
  }\href@noop {} {\bibfield  {journal} {\bibinfo  {journal} {Adv. Mater.}\
  }\textbf {\bibinfo {volume} {30}},\ \bibinfo {pages} {1705942} (\bibinfo
  {year} {2018})}\BibitemShut {NoStop}%
\bibitem [{\citenamefont {Wu}\ \emph {et~al.}(2019)\citenamefont {Wu},
  \citenamefont {Xie}, \citenamefont {Xu},\ and\ \citenamefont
  {He}}]{geteferroelectric}%
  \BibitemOpen
  \bibfield  {author} {\bibinfo {author} {\bibfnamefont {Di}~\bibnamefont
  {Wu}}, \bibinfo {author} {\bibfnamefont {Lin}\ \bibnamefont {Xie}}, \bibinfo
  {author} {\bibfnamefont {Xiao}\ \bibnamefont {Xu}}, \ and\ \bibinfo {author}
  {\bibfnamefont {Jiaqing}\ \bibnamefont {He}},\ }\bibfield  {title} {High
  Thermoelectric Performance Achieved in GeTe?Bi2Te3 Pseudo-Binary via Van der
  Waals Gap-Induced Hierarchical Ferroelectric Domain Structure,\ }\href@noop
  {} {\bibfield  {journal} {\bibinfo  {journal} {Adv. Funct. Mater.}\ }\textbf
  {\bibinfo {volume} {18}},\ \bibinfo {pages} {1806613} (\bibinfo {year}
  {2019})}\BibitemShut {NoStop}%
\bibitem [{\citenamefont {Li}\ \emph {et~al.}(2018)\citenamefont {Li},
  \citenamefont {Zhang}, \citenamefont {Chen}, \citenamefont {Lin},
  \citenamefont {Li}, \citenamefont {Shen}, \citenamefont {Witting},
  \citenamefont {Faghaninia}, \citenamefont {Chen}, \citenamefont {Jain},
  \citenamefont {Chen}, \citenamefont {Snyder},\ and\ \citenamefont
  {YanzhongPei}}]{getejoule}%
  \BibitemOpen
  \bibfield  {author} {\bibinfo {author} {\bibfnamefont {Juan}\ \bibnamefont
  {Li}}, \bibinfo {author} {\bibfnamefont {Xinyue}\ \bibnamefont {Zhang}},
  \bibinfo {author} {\bibfnamefont {Zhiwei}\ \bibnamefont {Chen}}, \bibinfo
  {author} {\bibfnamefont {Siqi}\ \bibnamefont {Lin}}, \bibinfo {author}
  {\bibfnamefont {Wen}\ \bibnamefont {Li}}, \bibinfo {author} {\bibfnamefont
  {Jiahong}\ \bibnamefont {Shen}}, \bibinfo {author} {\bibfnamefont {Ian~T.}\
  \bibnamefont {Witting}}, \bibinfo {author} {\bibfnamefont {Alireza}\
  \bibnamefont {Faghaninia}}, \bibinfo {author} {\bibfnamefont {Yue}\
  \bibnamefont {Chen}}, \bibinfo {author} {\bibfnamefont {Anubhav}\
  \bibnamefont {Jain}}, \bibinfo {author} {\bibfnamefont {Lidong}\ \bibnamefont
  {Chen}}, \bibinfo {author} {\bibfnamefont {G.~Jeffrey}\ \bibnamefont
  {Snyder}}, \ and\ \bibinfo {author} {\bibnamefont {YanzhongPei}},\ }\bibfield
   {title} {Low-Symmetry Rhombohedral GeTe Thermoelectrics,\ }\href@noop {}
  {\bibfield  {journal} {\bibinfo  {journal} {Joule}\ }\textbf {\bibinfo
  {volume} {2}},\ \bibinfo {pages} {976--987} (\bibinfo {year}
  {2018})}\BibitemShut {NoStop}%
\bibitem [{\citenamefont {Poudel}\ \emph {et~al.}(2008)\citenamefont {Poudel},
  \citenamefont {Hao}, \citenamefont {Ma}, \citenamefont {Lan}, \citenamefont
  {Minnich}, \citenamefont {Yu}, \citenamefont {Yang}, \citenamefont {Wang},
  \citenamefont {Muto}, \citenamefont {Vashaee}, \citenamefont {Chen},
  \citenamefont {Liu}, \citenamefont {Dresselhaus}, \citenamefont {Chen},\ and\
  \citenamefont {Ren}}]{bisbtenano}%
  \BibitemOpen
  \bibfield  {author} {\bibinfo {author} {\bibfnamefont {B.}~\bibnamefont
  {Poudel}}, \bibinfo {author} {\bibfnamefont {Q.}~\bibnamefont {Hao}},
  \bibinfo {author} {\bibfnamefont {Y.}~\bibnamefont {Ma}}, \bibinfo {author}
  {\bibfnamefont {Y.}~\bibnamefont {Lan}}, \bibinfo {author} {\bibfnamefont
  {A.}~\bibnamefont {Minnich}}, \bibinfo {author} {\bibfnamefont
  {B.}~\bibnamefont {Yu}}, \bibinfo {author} {\bibfnamefont {X.}~\bibnamefont
  {Yang}}, \bibinfo {author} {\bibfnamefont {D.}~\bibnamefont {Wang}}, \bibinfo
  {author} {\bibfnamefont {A.}~\bibnamefont {Muto}}, \bibinfo {author}
  {\bibfnamefont {D.}~\bibnamefont {Vashaee}}, \bibinfo {author} {\bibfnamefont
  {X.}~\bibnamefont {Chen}}, \bibinfo {author} {\bibfnamefont {J.}~\bibnamefont
  {Liu}}, \bibinfo {author} {\bibfnamefont {M.}~\bibnamefont {Dresselhaus}},
  \bibinfo {author} {\bibfnamefont {G.}~\bibnamefont {Chen}}, \ and\ \bibinfo
  {author} {\bibfnamefont {Z.}~\bibnamefont {Ren}},\ }\bibfield  {title}
  {High-Thermoelectric Performance of Nanostructured Bismuth Antimony Telluride
  Bulk Alloys,\ }\href@noop {} {\bibfield  {journal} {\bibinfo  {journal}
  {Science}\ }\textbf {\bibinfo {volume} {320}},\ \bibinfo {pages} {634--638}
  (\bibinfo {year} {2008})}\BibitemShut {NoStop}%
\bibitem [{\citenamefont {Kim}\ \emph {et~al.}(2015)\citenamefont {Kim},
  \citenamefont {Lee}, \citenamefont {Mun}, \citenamefont {Kim}, \citenamefont
  {Hwang}, \citenamefont {Roh}, \citenamefont {Yang}, \citenamefont {Shin},
  \citenamefont {Li}, \citenamefont {Lee}, \citenamefont {Snyder},\ and\
  \citenamefont {Kim}}]{bisbtedislocation}%
  \BibitemOpen
  \bibfield  {author} {\bibinfo {author} {\bibfnamefont {Sang~Il}\ \bibnamefont
  {Kim}}, \bibinfo {author} {\bibfnamefont {Kyu~Hyoung}\ \bibnamefont {Lee}},
  \bibinfo {author} {\bibfnamefont {Hyeon~A}\ \bibnamefont {Mun}}, \bibinfo
  {author} {\bibfnamefont {Hyun~Sik}\ \bibnamefont {Kim}}, \bibinfo {author}
  {\bibfnamefont {Sung~Woo}\ \bibnamefont {Hwang}}, \bibinfo {author}
  {\bibfnamefont {Jong~Wook}\ \bibnamefont {Roh}}, \bibinfo {author}
  {\bibfnamefont {Dae~Jin}\ \bibnamefont {Yang}}, \bibinfo {author}
  {\bibfnamefont {Weon~Ho}\ \bibnamefont {Shin}}, \bibinfo {author}
  {\bibfnamefont {Xiang~Shu}\ \bibnamefont {Li}}, \bibinfo {author}
  {\bibfnamefont {Young~Hee}\ \bibnamefont {Lee}}, \bibinfo {author}
  {\bibfnamefont {G.~Jeffrey}\ \bibnamefont {Snyder}}, \ and\ \bibinfo {author}
  {\bibfnamefont {Sung~Wng}\ \bibnamefont {Kim}},\ }\bibfield  {title} {Dense
  dislocation arrays embedded in grain boundaries for high-performance bulk
  thermoelectrics,\ }\href@noop {} {\bibfield  {journal} {\bibinfo  {journal}
  {Science}\ }\textbf {\bibinfo {volume} {348}},\ \bibinfo {pages} {109--114}
  (\bibinfo {year} {2015})}\BibitemShut {NoStop}%
\bibitem [{\citenamefont {Hu}\ \emph {et~al.}(2015)\citenamefont {Hu},
  \citenamefont {Wu}, \citenamefont {Zhu}, \citenamefont {Fu}, \citenamefont
  {He}, \citenamefont {Ying},\ and\ \citenamefont {Zhao}}]{bitese}%
  \BibitemOpen
  \bibfield  {author} {\bibinfo {author} {\bibfnamefont {L.}~\bibnamefont
  {Hu}}, \bibinfo {author} {\bibfnamefont {H.}~\bibnamefont {Wu}}, \bibinfo
  {author} {\bibfnamefont {T.}~\bibnamefont {Zhu}}, \bibinfo {author}
  {\bibfnamefont {C.}~\bibnamefont {Fu}}, \bibinfo {author} {\bibfnamefont
  {J.}~\bibnamefont {He}}, \bibinfo {author} {\bibfnamefont {P.}~\bibnamefont
  {Ying}}, \ and\ \bibinfo {author} {\bibfnamefont {X.}~\bibnamefont {Zhao}},\
  }\bibfield  {title} {Tuning multiscale microstructures to enhance
  thermoelectric performance of $n$-type bismuth-telluride-based solid
  solutions,\ }\href@noop {} {\bibfield  {journal} {\bibinfo  {journal} {Adv.
  Energy Mater.}\ }\textbf {\bibinfo {volume} {5}},\ \bibinfo {pages} {1500411}
  (\bibinfo {year} {2015})}\BibitemShut {NoStop}%
\bibitem [{\citenamefont {Zhang}\ \emph {et~al.}(2017)\citenamefont {Zhang},
  \citenamefont {Song}, \citenamefont {Pedersen}, \citenamefont {Yin},
  \citenamefont {Hung},\ and\ \citenamefont {Iversen}}]{mg3sb2discovery}%
  \BibitemOpen
  \bibfield  {author} {\bibinfo {author} {\bibfnamefont {Jiawei}\ \bibnamefont
  {Zhang}}, \bibinfo {author} {\bibfnamefont {Lirong}\ \bibnamefont {Song}},
  \bibinfo {author} {\bibfnamefont {Steffen~Hindborg}\ \bibnamefont
  {Pedersen}}, \bibinfo {author} {\bibfnamefont {Hao}\ \bibnamefont {Yin}},
  \bibinfo {author} {\bibfnamefont {Le~Thanh}\ \bibnamefont {Hung}}, \ and\
  \bibinfo {author} {\bibfnamefont {Bo~Brummerstedt}\ \bibnamefont {Iversen}},\
  }\bibfield  {title} {Discovery of high-performance low-cost $n$-type
  Mg$_{3}$Sb$_{2}$-based thermoelectric materials with multi-valley conduction
  bands,\ }\href@noop {} {\bibfield  {journal} {\bibinfo  {journal} {Nat.
  Commun.}\ }\textbf {\bibinfo {volume} {8}} (\bibinfo {year}
  {2017})}\BibitemShut {NoStop}%
\bibitem [{\citenamefont {Wood}\ \emph {et~al.}(2019)\citenamefont {Wood},
  \citenamefont {Kuo}, \citenamefont {Imasato},\ and\ \citenamefont
  {Snyder}}]{lowtmg3sb2}%
  \BibitemOpen
  \bibfield  {author} {\bibinfo {author} {\bibfnamefont {Maxwell}\ \bibnamefont
  {Wood}}, \bibinfo {author} {\bibfnamefont {Jimmy~Jiahong}\ \bibnamefont
  {Kuo}}, \bibinfo {author} {\bibfnamefont {Kazuki}\ \bibnamefont {Imasato}}, \
  and\ \bibinfo {author} {\bibfnamefont {Gerald~Jeffrey}\ \bibnamefont
  {Snyder}},\ }\bibfield  {title} {Improvement of Low-Temperature $zT$ in a
  Mg$_{3}$Sb$_{2}$?Mg$_{3}$Bi$_{2}$ Solid Solution via Mg-Vapor Annealing,\
  }\href@noop {} {\bibfield  {journal} {\bibinfo  {journal} {Adv. Mater.}\
  }\textbf {\bibinfo {volume} {31}} (\bibinfo {year} {2019})}\BibitemShut
  {NoStop}%
\bibitem [{\citenamefont {Shi}\ \emph {et~al.}(2019)\citenamefont {Shi},
  \citenamefont {Sun}, \citenamefont {Bu}, \citenamefont {Zhang}, \citenamefont
  {Wu}, \citenamefont {Lin}, \citenamefont {Li}, \citenamefont {Faghaninia},
  \citenamefont {Jain},\ and\ \citenamefont {Pei}}]{mg3sb2revelation}%
  \BibitemOpen
  \bibfield  {author} {\bibinfo {author} {\bibfnamefont {Xuemin}\ \bibnamefont
  {Shi}}, \bibinfo {author} {\bibfnamefont {Cheng}\ \bibnamefont {Sun}},
  \bibinfo {author} {\bibfnamefont {Zhonglin}\ \bibnamefont {Bu}}, \bibinfo
  {author} {\bibfnamefont {Xinyue}\ \bibnamefont {Zhang}}, \bibinfo {author}
  {\bibfnamefont {Yixuan}\ \bibnamefont {Wu}}, \bibinfo {author} {\bibfnamefont
  {Siqi}\ \bibnamefont {Lin}}, \bibinfo {author} {\bibfnamefont {Wen}\
  \bibnamefont {Li}}, \bibinfo {author} {\bibfnamefont {Alireza}\ \bibnamefont
  {Faghaninia}}, \bibinfo {author} {\bibfnamefont {Anubhav}\ \bibnamefont
  {Jain}}, \ and\ \bibinfo {author} {\bibfnamefont {Yanzhong}\ \bibnamefont
  {Pei}},\ }\bibfield  {title} {Revelation of Inherently High Mobility Enables
  Mg$_{3}$Sb$_{2}$ as a Sustainable Alternative to $n$-Bi$_{2}$Te$_{3}$
  Thermoelectrics,\ }\href@noop {} {\bibfield  {journal} {\bibinfo  {journal}
  {Adv. Sci.}\ }\textbf {\bibinfo {volume} {6}} (\bibinfo {year}
  {2019})}\BibitemShut {NoStop}%
\bibitem [{\citenamefont {Mao}\ \emph {et~al.}(2019)\citenamefont {Mao},
  \citenamefont {Zhu}, \citenamefont {Ding}, \citenamefont {Liu}, \citenamefont
  {Amila}, \citenamefont {Gamage}, \citenamefont {Chen},\ and\ \citenamefont
  {Ren}}]{mg3bi2cooling}%
  \BibitemOpen
  \bibfield  {author} {\bibinfo {author} {\bibfnamefont {Jun}\ \bibnamefont
  {Mao}}, \bibinfo {author} {\bibfnamefont {Hangtian}\ \bibnamefont {Zhu}},
  \bibinfo {author} {\bibfnamefont {Zhiwei}\ \bibnamefont {Ding}}, \bibinfo
  {author} {\bibfnamefont {Zihang}\ \bibnamefont {Liu}}, \bibinfo {author}
  {\bibfnamefont {Geethal}\ \bibnamefont {Amila}}, \bibinfo {author}
  {\bibnamefont {Gamage}}, \bibinfo {author} {\bibfnamefont {Gang}\
  \bibnamefont {Chen}}, \ and\ \bibinfo {author} {\bibfnamefont {Zhifeng}\
  \bibnamefont {Ren}},\ }\bibfield  {title} {High thermoelectric cooling
  performance of $n$-type Mg$_{3}$Bi$_{2}$-based materials,\ }\href@noop {}
  {\bibfield  {journal} {\bibinfo  {journal} {Science}\ }\textbf {\bibinfo
  {volume} {365}} (\bibinfo {year} {2019})}\BibitemShut {NoStop}%
\bibitem [{\citenamefont {Bell}(2008)}]{thermoelectricsystem}%
  \BibitemOpen
  \bibfield  {author} {\bibinfo {author} {\bibfnamefont {Lon~E.}\ \bibnamefont
  {Bell}},\ }\bibfield  {title} {Cooling, Heating, Generating Power, and
  Recovering Waste Heat with Thermoelectric Systems,\ }\href@noop {} {\bibfield
   {journal} {\bibinfo  {journal} {Science}\ }\textbf {\bibinfo {volume}
  {321}},\ \bibinfo {pages} {1457--1461} (\bibinfo {year} {2008})}\BibitemShut
  {NoStop}%
\bibitem [{\citenamefont {He}\ and\ \citenamefont
  {Tritt}(2017)}]{advancesinthermoelectricmaterials}%
  \BibitemOpen
  \bibfield  {author} {\bibinfo {author} {\bibfnamefont {Jian}\ \bibnamefont
  {He}}\ and\ \bibinfo {author} {\bibfnamefont {Terry~M.}\ \bibnamefont
  {Tritt}},\ }\bibfield  {title} {Advances in Thermoelectric Materials
  Research: Looking back and moving forward,\ }\href@noop {} {\bibfield
  {journal} {\bibinfo  {journal} {Science}\ }\textbf {\bibinfo {volume}
  {357}},\ \bibinfo {pages} {1--9} (\bibinfo {year} {2017})}\BibitemShut
  {NoStop}%
\bibitem [{\citenamefont {Tritt}\ and\ \citenamefont
  {Subramanian}(2006)}]{birdeye}%
  \BibitemOpen
  \bibfield  {author} {\bibinfo {author} {\bibfnamefont {T.M.}\ \bibnamefont
  {Tritt}}\ and\ \bibinfo {author} {\bibfnamefont {M.A.}\ \bibnamefont
  {Subramanian}},\ }\bibfield  {title} {Thermoelectric materials, phenomena,
  and applications: a birdÕs eye view,\ }\href@noop {} {\bibfield  {journal}
  {\bibinfo  {journal} {MRS Bulletin}\ }\textbf {\bibinfo {volume} {31}},\
  \bibinfo {pages} {188--198} (\bibinfo {year} {2006})}\BibitemShut {NoStop}%
\bibitem [{\citenamefont {Yang}\ and\ \citenamefont
  {Caillat}(2006)}]{spaceauto}%
  \BibitemOpen
  \bibfield  {author} {\bibinfo {author} {\bibfnamefont {J.}~\bibnamefont
  {Yang}}\ and\ \bibinfo {author} {\bibfnamefont {T.}~\bibnamefont {Caillat}},\
  }\bibfield  {title} {Thermoelectric materials for space and automotive power
  generation,\ }\href@noop {} {\bibfield  {journal} {\bibinfo  {journal} {MRS
  Bulletin}\ }\textbf {\bibinfo {volume} {31}},\ \bibinfo {pages} {224--229}
  (\bibinfo {year} {2006})}\BibitemShut {NoStop}%
\bibitem [{\citenamefont {Snyder}\ and\ \citenamefont
  {Toberer}(2008)}]{complex}%
  \BibitemOpen
  \bibfield  {author} {\bibinfo {author} {\bibfnamefont {J.}~\bibnamefont
  {Snyder}}\ and\ \bibinfo {author} {\bibfnamefont {E.}~\bibnamefont
  {Toberer}},\ }\bibfield  {title} {Complex thermoelectric materials,\
  }\href@noop {} {\bibfield  {journal} {\bibinfo  {journal} {Nat. Mater.}\
  }\textbf {\bibinfo {volume} {7}},\ \bibinfo {pages} {105--114} (\bibinfo
  {year} {2008})}\BibitemShut {NoStop}%
\bibitem [{\citenamefont {Sootsman}\ \emph {et~al.}(2009)\citenamefont
  {Sootsman}, \citenamefont {Chung},\ and\ \citenamefont
  {Kanatzidis}}]{newandold}%
  \BibitemOpen
  \bibfield  {author} {\bibinfo {author} {\bibfnamefont {J.~R.}\ \bibnamefont
  {Sootsman}}, \bibinfo {author} {\bibfnamefont {D.}~\bibnamefont {Chung}}, \
  and\ \bibinfo {author} {\bibfnamefont {M.G.}\ \bibnamefont {Kanatzidis}},\
  }\bibfield  {title} {New and old concepts in thermoelectric materials,\
  }\href@noop {} {\bibfield  {journal} {\bibinfo  {journal} {Angew. Chem.}\
  }\textbf {\bibinfo {volume} {48}},\ \bibinfo {pages} {8616--8639} (\bibinfo
  {year} {2009})}\BibitemShut {NoStop}%
\bibitem [{\citenamefont {Zeier}\ \emph {et~al.}(2016)\citenamefont {Zeier},
  \citenamefont {Zevalkink}, \citenamefont {Gibbs}, \citenamefont {Hautier},
  \citenamefont {Kanatzidis},\ and\ \citenamefont {Snyder}}]{intuition}%
  \BibitemOpen
  \bibfield  {author} {\bibinfo {author} {\bibfnamefont {W.~G.}\ \bibnamefont
  {Zeier}}, \bibinfo {author} {\bibfnamefont {A.}~\bibnamefont {Zevalkink}},
  \bibinfo {author} {\bibfnamefont {Z.~M.}\ \bibnamefont {Gibbs}}, \bibinfo
  {author} {\bibfnamefont {G.}~\bibnamefont {Hautier}}, \bibinfo {author}
  {\bibfnamefont {M.~G.}\ \bibnamefont {Kanatzidis}}, \ and\ \bibinfo {author}
  {\bibfnamefont {G.~J.}\ \bibnamefont {Snyder}},\ }\bibfield  {title}
  {Thinking Like a Chemist: Intuition in Thermoelectric Materials,\ }\href@noop
  {} {\bibfield  {journal} {\bibinfo  {journal} {Angew. Chem.}\ }\textbf
  {\bibinfo {volume} {55}},\ \bibinfo {pages} {6826--6841} (\bibinfo {year}
  {2016})}\BibitemShut {NoStop}%
\bibitem [{\citenamefont {Xiao~Zhang}(2015)}]{thermoelectricmaterials}%
  \BibitemOpen
  \bibfield  {author} {\bibinfo {author} {\bibfnamefont {Li-Dong~Zhao}\
  \bibnamefont {Xiao~Zhang}},\ }\bibfield  {title} {Thermoelectric materials:
  Energy conversion between heat and electricity,\ }\href@noop {} {\bibfield
  {journal} {\bibinfo  {journal} {J. Materiomics.}\ }\textbf {\bibinfo {volume}
  {1}},\ \bibinfo {pages} {92--105} (\bibinfo {year} {2015})}\BibitemShut
  {NoStop}%
\bibitem [{\citenamefont {Zhu}\ \emph {et~al.}(2017)\citenamefont {Zhu},
  \citenamefont {Liu}, \citenamefont {Fu}, \citenamefont {Heremans},
  \citenamefont {Snyder},\ and\ \citenamefont {Zhao}}]{compromisesynergy}%
  \BibitemOpen
  \bibfield  {author} {\bibinfo {author} {\bibfnamefont {T.}~\bibnamefont
  {Zhu}}, \bibinfo {author} {\bibfnamefont {Y.}~\bibnamefont {Liu}}, \bibinfo
  {author} {\bibfnamefont {C.}~\bibnamefont {Fu}}, \bibinfo {author}
  {\bibfnamefont {J.~P.}\ \bibnamefont {Heremans}}, \bibinfo {author}
  {\bibfnamefont {J.~G.}\ \bibnamefont {Snyder}}, \ and\ \bibinfo {author}
  {\bibfnamefont {X.}~\bibnamefont {Zhao}},\ }\bibfield  {title} {Compromise
  and Synergy in High-Efficiency Thermoelectric Materials,\ }\href@noop {}
  {\bibfield  {journal} {\bibinfo  {journal} {Adv. Mater.}\ }\textbf {\bibinfo
  {volume} {29}},\ \bibinfo {pages} {1605884} (\bibinfo {year}
  {2017})}\BibitemShut {NoStop}%
\bibitem [{\citenamefont {Mao}\ \emph {et~al.}(2018)\citenamefont {Mao},
  \citenamefont {Liu}, \citenamefont {Zhou}, \citenamefont {Zhu}, \citenamefont
  {Zhang}, \citenamefont {Chen},\ and\ \citenamefont
  {Ren}}]{advancesinthermoelectrics}%
  \BibitemOpen
  \bibfield  {author} {\bibinfo {author} {\bibfnamefont {Jun}\ \bibnamefont
  {Mao}}, \bibinfo {author} {\bibfnamefont {Zihang}\ \bibnamefont {Liu}},
  \bibinfo {author} {\bibfnamefont {Jiawei}\ \bibnamefont {Zhou}}, \bibinfo
  {author} {\bibfnamefont {Hangtian}\ \bibnamefont {Zhu}}, \bibinfo {author}
  {\bibfnamefont {Qian}\ \bibnamefont {Zhang}}, \bibinfo {author}
  {\bibfnamefont {Gang}\ \bibnamefont {Chen}}, \ and\ \bibinfo {author}
  {\bibfnamefont {Zhifeng}\ \bibnamefont {Ren}},\ }\bibfield  {title} {Advances
  in Thermoelectrics,\ }\href@noop {} {\bibfield  {journal} {\bibinfo
  {journal} {Adv. in Phys.}\ }\textbf {\bibinfo {volume} {67}},\ \bibinfo
  {pages} {69--147} (\bibinfo {year} {2018})}\BibitemShut {NoStop}%
\bibitem [{\citenamefont {Pei}\ \emph {et~al.}(2012{\natexlab{a}})\citenamefont
  {Pei}, \citenamefont {LaLonde}, \citenamefont {Wang},\ and\ \citenamefont
  {Snyder}}]{loweffmass}%
  \BibitemOpen
  \bibfield  {author} {\bibinfo {author} {\bibfnamefont {Yanzhong}\
  \bibnamefont {Pei}}, \bibinfo {author} {\bibfnamefont {Aaron~D.}\
  \bibnamefont {LaLonde}}, \bibinfo {author} {\bibfnamefont {Heng}\
  \bibnamefont {Wang}}, \ and\ \bibinfo {author} {\bibfnamefont {G.~Jeffrey}\
  \bibnamefont {Snyder}},\ }\bibfield  {title} {Low Effective Mass Leading to
  High Thermoelectric Performance,\ }\href@noop {} {\bibfield  {journal}
  {\bibinfo  {journal} {Energy Environ. Sci.}\ }\textbf {\bibinfo {volume}
  {5}},\ \bibinfo {pages} {7963--7969} (\bibinfo {year}
  {2012}{\natexlab{a}})}\BibitemShut {NoStop}%
\bibitem [{\citenamefont {Gorai}\ \emph {et~al.}(2017)\citenamefont {Gorai},
  \citenamefont {Stevanovi{\'c}},\ and\ \citenamefont
  {Toberer}}]{computationalthermoelectrics}%
  \BibitemOpen
  \bibfield  {author} {\bibinfo {author} {\bibfnamefont {Prashun}\ \bibnamefont
  {Gorai}}, \bibinfo {author} {\bibfnamefont {Vladan}\ \bibnamefont
  {Stevanovi{\'c}}}, \ and\ \bibinfo {author} {\bibfnamefont {Eric~S.}\
  \bibnamefont {Toberer}},\ }\bibfield  {title} {Computationally guided
  discovery of thermoelectric materials,\ }\href@noop {} {\bibfield  {journal}
  {\bibinfo  {journal} {Nat. Rev. Mater.}\ }\textbf {\bibinfo {volume} {2}},\
  \bibinfo {pages} {1--16} (\bibinfo {year} {2017})}\BibitemShut {NoStop}%
\bibitem [{\citenamefont {Fu}\ \emph {et~al.}(2014)\citenamefont {Fu},
  \citenamefont {Zhu}, \citenamefont {Pei}, \citenamefont {Xie}, \citenamefont
  {Wang}, \citenamefont {Snyder}, \citenamefont {Liu}, \citenamefont {Liu},\
  and\ \citenamefont {Zhao}}]{halfheuslerbanddegeneracy}%
  \BibitemOpen
  \bibfield  {author} {\bibinfo {author} {\bibfnamefont {Chenguang}\
  \bibnamefont {Fu}}, \bibinfo {author} {\bibfnamefont {Tiejun}\ \bibnamefont
  {Zhu}}, \bibinfo {author} {\bibfnamefont {Yanzhong}\ \bibnamefont {Pei}},
  \bibinfo {author} {\bibfnamefont {Hanhui}\ \bibnamefont {Xie}}, \bibinfo
  {author} {\bibfnamefont {Heng}\ \bibnamefont {Wang}}, \bibinfo {author}
  {\bibfnamefont {G.~Jeffrey}\ \bibnamefont {Snyder}}, \bibinfo {author}
  {\bibfnamefont {Yong}\ \bibnamefont {Liu}}, \bibinfo {author} {\bibfnamefont
  {Yintu}\ \bibnamefont {Liu}}, \ and\ \bibinfo {author} {\bibfnamefont
  {Xinbing}\ \bibnamefont {Zhao}},\ }\bibfield  {title} {High Band Degeneracy
  Contributes to High Thermoelectric Performance in $p$-Type Half-Heusler
  Compounds,\ }\href@noop {} {\bibfield  {journal} {\bibinfo  {journal} {Adv.
  Energy Mater.}\ }\textbf {\bibinfo {volume} {4}},\ \bibinfo {pages} {1400600}
  (\bibinfo {year} {2014})}\BibitemShut {NoStop}%
\bibitem [{\citenamefont {Gibbs}\ \emph {et~al.}(2017)\citenamefont {Gibbs},
  \citenamefont {Ricci}, \citenamefont {Li}, \citenamefont {Zhu}, \citenamefont
  {Persson}, \citenamefont {Ceder}, \citenamefont {Hautier}, \citenamefont
  {Jain},\ and\ \citenamefont {Snyder}}]{complexityfactor}%
  \BibitemOpen
  \bibfield  {author} {\bibinfo {author} {\bibfnamefont {Zachary~M.}\
  \bibnamefont {Gibbs}}, \bibinfo {author} {\bibfnamefont {Francesco}\
  \bibnamefont {Ricci}}, \bibinfo {author} {\bibfnamefont {Guodong}\
  \bibnamefont {Li}}, \bibinfo {author} {\bibfnamefont {Hong}\ \bibnamefont
  {Zhu}}, \bibinfo {author} {\bibfnamefont {Kristin}\ \bibnamefont {Persson}},
  \bibinfo {author} {\bibfnamefont {Gerbrand}\ \bibnamefont {Ceder}}, \bibinfo
  {author} {\bibfnamefont {Geoffroy}\ \bibnamefont {Hautier}}, \bibinfo
  {author} {\bibfnamefont {Anubhav}\ \bibnamefont {Jain}}, \ and\ \bibinfo
  {author} {\bibfnamefont {G.~Jeffrey}\ \bibnamefont {Snyder}},\ }\bibfield
  {title} {Effective mass and Fermi surface complexity factor from ab initio
  band structure calculations,\ }\href@noop {} {\bibfield  {journal} {\bibinfo
  {journal} {Comput. Mater.}\ }\textbf {\bibinfo {volume} {3}},\ \bibinfo
  {pages} {1--7} (\bibinfo {year} {2017})}\BibitemShut {NoStop}%
\bibitem [{\citenamefont {Yan}\ \emph {et~al.}(2015)\citenamefont {Yan},
  \citenamefont {Gorai}, \citenamefont {Ortiz}, \citenamefont {Miller},
  \citenamefont {Barnett}, \citenamefont {Mason}, \citenamefont
  {Stevanovi{\'c}},\ and\ \citenamefont {Toberer}}]{materialdescriptors}%
  \BibitemOpen
  \bibfield  {author} {\bibinfo {author} {\bibfnamefont {Jun}\ \bibnamefont
  {Yan}}, \bibinfo {author} {\bibfnamefont {Prashun}\ \bibnamefont {Gorai}},
  \bibinfo {author} {\bibfnamefont {Brenden}\ \bibnamefont {Ortiz}}, \bibinfo
  {author} {\bibfnamefont {Sam}\ \bibnamefont {Miller}}, \bibinfo {author}
  {\bibfnamefont {Scott~A.}\ \bibnamefont {Barnett}}, \bibinfo {author}
  {\bibfnamefont {Thomas}\ \bibnamefont {Mason}}, \bibinfo {author}
  {\bibfnamefont {Vladan}\ \bibnamefont {Stevanovi{\'c}}}, \ and\ \bibinfo
  {author} {\bibfnamefont {Eric~S.}\ \bibnamefont {Toberer}},\ }\bibfield
  {title} {Material descriptors for predicting thermoelectric performance,\
  }\href@noop {} {\bibfield  {journal} {\bibinfo  {journal} {Energy Environ.
  Sci.}\ }\textbf {\bibinfo {volume} {8}},\ \bibinfo {pages} {983--994}
  (\bibinfo {year} {2015})}\BibitemShut {NoStop}%
\bibitem [{\citenamefont {Park}\ \emph {et~al.}(2019)\citenamefont {Park},
  \citenamefont {Xia},\ and\ \citenamefont {Ozoli\c{n}\v{s}}}]{ba2biau}%
  \BibitemOpen
  \bibfield  {author} {\bibinfo {author} {\bibfnamefont {Junsoo}\ \bibnamefont
  {Park}}, \bibinfo {author} {\bibfnamefont {Yi}~\bibnamefont {Xia}}, \ and\
  \bibinfo {author} {\bibfnamefont {Vidvuds}\ \bibnamefont {Ozoli\c{n}\v{s}}},\
  }\bibfield  {title} {High Thermoelectric Power Factor and Efficiency from a
  Highly Dispersive Band in Ba$_{2}$BiAu,\ }\href@noop {} {\bibfield  {journal}
  {\bibinfo  {journal} {Phys. Rev. Appl.}\ }\textbf {\bibinfo {volume} {11}},\
  \bibinfo {pages} {014058} (\bibinfo {year} {2019})}\BibitemShut {NoStop}%
\bibitem [{\citenamefont {Ma}\ \emph {et~al.}(2020)\citenamefont {Ma},
  \citenamefont {Nissimagoudar}, \citenamefont {Wang},\ and\ \citenamefont
  {Li}}]{ba2biauredo}%
  \BibitemOpen
  \bibfield  {author} {\bibinfo {author} {\bibfnamefont {Jinlong}\ \bibnamefont
  {Ma}}, \bibinfo {author} {\bibfnamefont {Arun~S.}\ \bibnamefont
  {Nissimagoudar}}, \bibinfo {author} {\bibfnamefont {Shudong}\ \bibnamefont
  {Wang}}, \ and\ \bibinfo {author} {\bibfnamefont {Wu}~\bibnamefont {Li}},\
  }\bibfield  {title} {High Thermoelectric Figure of Merit of Full-Heusler
  Ba$_{2}$AuX (X = As, Sb, and Bi),\ }\href@noop {} {\bibfield  {journal}
  {\bibinfo  {journal} {Phys. Status Solidi RRL}\ }\textbf {\bibinfo {volume}
  {14}},\ \bibinfo {pages} {2000084} (\bibinfo {year} {2020})}\BibitemShut
  {NoStop}%
\bibitem [{\citenamefont {He}\ \emph {et~al.}(2016{\natexlab{a}})\citenamefont
  {He}, \citenamefont {Amsler}, \citenamefont {Xia}, \citenamefont {Naghavi},
  \citenamefont {Hegde}, \citenamefont {Hao}, \citenamefont {Goedecker},
  \citenamefont {Ozoli\c{n}\v{s}},\ and\ \citenamefont
  {Wolverton}}]{ultralowheusler}%
  \BibitemOpen
  \bibfield  {author} {\bibinfo {author} {\bibfnamefont {J.}~\bibnamefont
  {He}}, \bibinfo {author} {\bibfnamefont {M.}~\bibnamefont {Amsler}}, \bibinfo
  {author} {\bibfnamefont {Y.}~\bibnamefont {Xia}}, \bibinfo {author}
  {\bibfnamefont {S.~S.}\ \bibnamefont {Naghavi}}, \bibinfo {author}
  {\bibfnamefont {V.}~\bibnamefont {Hegde}}, \bibinfo {author} {\bibfnamefont
  {S.}~\bibnamefont {Hao}}, \bibinfo {author} {\bibfnamefont {S.}~\bibnamefont
  {Goedecker}}, \bibinfo {author} {\bibfnamefont {V.}~\bibnamefont
  {Ozoli\c{n}\v{s}}}, \ and\ \bibinfo {author} {\bibfnamefont {C.}~\bibnamefont
  {Wolverton}},\ }\bibfield  {title} {Ultralow Thermal Conductivity in Full
  Heusler Semiconductors,\ }\href@noop {} {\bibfield  {journal} {\bibinfo
  {journal} {Phys. Rev. Lett.}\ }\textbf {\bibinfo {volume} {117}},\ \bibinfo
  {pages} {046602} (\bibinfo {year} {2016}{\natexlab{a}})}\BibitemShut
  {NoStop}%
\bibitem [{\citenamefont {Giannozzi}\ \emph {et~al.}(2009)\citenamefont
  {Giannozzi}, \citenamefont {Baroni}, \citenamefont {Bonini}, \citenamefont
  {Calandra}, \citenamefont {Car}, \citenamefont {Cavazzoni}, \citenamefont
  {Ceresoli}, \citenamefont {Chiarotti}, \citenamefont {Cococcioni},
  \citenamefont {Dabo}, \citenamefont {{Dal Corso}}, \citenamefont
  {de~Gironcoli}, \citenamefont {Fabris}, \citenamefont {Fratesi},
  \citenamefont {Gebauer}, \citenamefont {Gerstmann}, \citenamefont
  {Gougoussis}, \citenamefont {Kokalj}, \citenamefont {Lazzeri}, \citenamefont
  {Martin-Samos}, \citenamefont {Marzari}, \citenamefont {Mauri}, \citenamefont
  {Mazzarello}, \citenamefont {Paolini}, \citenamefont {Pasquarello},
  \citenamefont {Paulatto}, \citenamefont {Sbraccia}, \citenamefont {Scandolo},
  \citenamefont {Sclauzero}, \citenamefont {Seitsonen}, \citenamefont
  {Smogunov}, \citenamefont {Umari},\ and\ \citenamefont
  {Wentzcovitch}}]{qespresso1}%
  \BibitemOpen
  \bibfield  {author} {\bibinfo {author} {\bibfnamefont {Paolo}\ \bibnamefont
  {Giannozzi}}, \bibinfo {author} {\bibfnamefont {Stefano}\ \bibnamefont
  {Baroni}}, \bibinfo {author} {\bibfnamefont {Nicola}\ \bibnamefont {Bonini}},
  \bibinfo {author} {\bibfnamefont {Matteo}\ \bibnamefont {Calandra}}, \bibinfo
  {author} {\bibfnamefont {Roberto}\ \bibnamefont {Car}}, \bibinfo {author}
  {\bibfnamefont {Carlo}\ \bibnamefont {Cavazzoni}}, \bibinfo {author}
  {\bibfnamefont {Davide}\ \bibnamefont {Ceresoli}}, \bibinfo {author}
  {\bibfnamefont {Guido~L}\ \bibnamefont {Chiarotti}}, \bibinfo {author}
  {\bibfnamefont {Matteo}\ \bibnamefont {Cococcioni}}, \bibinfo {author}
  {\bibfnamefont {Ismaila}\ \bibnamefont {Dabo}}, \bibinfo {author}
  {\bibfnamefont {Andrea}\ \bibnamefont {{Dal Corso}}}, \bibinfo {author}
  {\bibfnamefont {Stefano}\ \bibnamefont {de~Gironcoli}}, \bibinfo {author}
  {\bibfnamefont {Stefano}\ \bibnamefont {Fabris}}, \bibinfo {author}
  {\bibfnamefont {Guido}\ \bibnamefont {Fratesi}}, \bibinfo {author}
  {\bibfnamefont {Ralph}\ \bibnamefont {Gebauer}}, \bibinfo {author}
  {\bibfnamefont {Uwe}\ \bibnamefont {Gerstmann}}, \bibinfo {author}
  {\bibfnamefont {Christos}\ \bibnamefont {Gougoussis}}, \bibinfo {author}
  {\bibfnamefont {Anton}\ \bibnamefont {Kokalj}}, \bibinfo {author}
  {\bibfnamefont {Michele}\ \bibnamefont {Lazzeri}}, \bibinfo {author}
  {\bibfnamefont {Layla}\ \bibnamefont {Martin-Samos}}, \bibinfo {author}
  {\bibfnamefont {Nicola}\ \bibnamefont {Marzari}}, \bibinfo {author}
  {\bibfnamefont {Francesco}\ \bibnamefont {Mauri}}, \bibinfo {author}
  {\bibfnamefont {Riccardo}\ \bibnamefont {Mazzarello}}, \bibinfo {author}
  {\bibfnamefont {Stefano}\ \bibnamefont {Paolini}}, \bibinfo {author}
  {\bibfnamefont {Alfredo}\ \bibnamefont {Pasquarello}}, \bibinfo {author}
  {\bibfnamefont {Lorenzo}\ \bibnamefont {Paulatto}}, \bibinfo {author}
  {\bibfnamefont {Carlo}\ \bibnamefont {Sbraccia}}, \bibinfo {author}
  {\bibfnamefont {Sandro}\ \bibnamefont {Scandolo}}, \bibinfo {author}
  {\bibfnamefont {Gabriele}\ \bibnamefont {Sclauzero}}, \bibinfo {author}
  {\bibfnamefont {Ari~P}\ \bibnamefont {Seitsonen}}, \bibinfo {author}
  {\bibfnamefont {Alexander}\ \bibnamefont {Smogunov}}, \bibinfo {author}
  {\bibfnamefont {Paolo}\ \bibnamefont {Umari}}, \ and\ \bibinfo {author}
  {\bibfnamefont {Renata~M}\ \bibnamefont {Wentzcovitch}},\ }\bibfield  {title}
  {QUANTUM ESPRESSO: a modular and open-source software project for quantum
  simulations of materials,\ }\href@noop {} {\bibfield  {journal} {\bibinfo
  {journal} {J. Phys. Condens. Matter}\ ,\ \bibinfo {pages} {395502 (19pp)}}
  (\bibinfo {year} {2009})}\BibitemShut {NoStop}%
\bibitem [{\citenamefont {Giannozzi}\ \emph {et~al.}(2017)\citenamefont
  {Giannozzi}, \citenamefont {Andreussi}, \citenamefont {Brumme}, \citenamefont
  {Bunau}, \citenamefont {Nardelli}, \citenamefont {Calandra}, \citenamefont
  {Car}, \citenamefont {Cavazzoni}, \citenamefont {Ceresoli},\ and\
  \citenamefont {Cococcioni}}]{qespresso2}%
  \BibitemOpen
  \bibfield  {author} {\bibinfo {author} {\bibfnamefont {P.}~\bibnamefont
  {Giannozzi}}, \bibinfo {author} {\bibfnamefont {O.}~\bibnamefont
  {Andreussi}}, \bibinfo {author} {\bibfnamefont {T.}~\bibnamefont {Brumme}},
  \bibinfo {author} {\bibfnamefont {O.}~\bibnamefont {Bunau}}, \bibinfo
  {author} {\bibfnamefont {M~Buongiorno}\ \bibnamefont {Nardelli}}, \bibinfo
  {author} {\bibfnamefont {M}~\bibnamefont {Calandra}}, \bibinfo {author}
  {\bibfnamefont {R.}~\bibnamefont {Car}}, \bibinfo {author} {\bibfnamefont
  {C.}~\bibnamefont {Cavazzoni}}, \bibinfo {author} {\bibfnamefont
  {D.}~\bibnamefont {Ceresoli}}, \ and\ \bibinfo {author} {\bibfnamefont
  {M.}~\bibnamefont {Cococcioni}},\ }\bibfield  {title} {Advanced capabilities
  for materials modeling with Quantum ESPRESSO,\ }\href@noop {} {\bibfield
  {journal} {\bibinfo  {journal} {J. Phys. Condens. Matter}\ ,\ \bibinfo
  {pages} {465901 (31pp)}} (\bibinfo {year} {2017})}\BibitemShut {NoStop}%
\bibitem [{\citenamefont {Hamann}(2013)}]{oncv1}%
  \BibitemOpen
  \bibfield  {author} {\bibinfo {author} {\bibfnamefont {D.~R.}\ \bibnamefont
  {Hamann}},\ }\bibfield  {title} {Optimized norm-conserving Vanderbilt
  pseudopotentials,\ }\href@noop {} {\bibfield  {journal} {\bibinfo  {journal}
  {Phys. Rev. B}\ }\textbf {\bibinfo {volume} {88}},\ \bibinfo {pages} {085117}
  (\bibinfo {year} {2013})}\BibitemShut {NoStop}%
\bibitem [{\citenamefont {Schlipf}\ and\ \citenamefont {Gygi}(2015)}]{oncv2}%
  \BibitemOpen
  \bibfield  {author} {\bibinfo {author} {\bibfnamefont {Martin}\ \bibnamefont
  {Schlipf}}\ and\ \bibinfo {author} {\bibfnamefont {Fran{\c c}ois}\
  \bibnamefont {Gygi}},\ }\bibfield  {title} {Optimization algorithm for the
  generation of \{ONCV\} pseudopotentials,\ }\href@noop {} {\bibfield
  {journal} {\bibinfo  {journal} {Comput. Phys. Commun.}\ }\textbf {\bibinfo
  {volume} {196}},\ \bibinfo {pages} {36--44} (\bibinfo {year}
  {2015})}\BibitemShut {NoStop}%
\bibitem [{\citenamefont {abd Marco~Govoni}\ \emph {et~al.}(2016)\citenamefont
  {abd Marco~Govoni}, \citenamefont {Hamada},\ and\ \citenamefont
  {Galli}}]{oncv3}%
  \BibitemOpen
  \bibfield  {author} {\bibinfo {author} {\bibfnamefont {Peter~Scherpel}\
  \bibnamefont {abd Marco~Govoni}}, \bibinfo {author} {\bibfnamefont {Ikutaro}\
  \bibnamefont {Hamada}}, \ and\ \bibinfo {author} {\bibfnamefont {Giulia}\
  \bibnamefont {Galli}},\ }\bibfield  {title} {Implementation and Validation of
  Fully Relativistic \textit{GW} Calculations: Spin?Orbit Coupling in
  Molecules, Nanocrystals, and Solids,\ }\href@noop {} {\bibfield  {journal}
  {\bibinfo  {journal} {J. Chem. Theory Comput.}\ }\textbf {\bibinfo {volume}
  {12}},\ \bibinfo {pages} {3523--3544} (\bibinfo {year} {2016})}\BibitemShut
  {NoStop}%
\bibitem [{\citenamefont {Perdew}\ \emph {et~al.}(1996)\citenamefont {Perdew},
  \citenamefont {Burke},\ and\ \citenamefont {Ernzerhof}}]{pbe}%
  \BibitemOpen
  \bibfield  {author} {\bibinfo {author} {\bibfnamefont {John~P.}\ \bibnamefont
  {Perdew}}, \bibinfo {author} {\bibfnamefont {Kieron}\ \bibnamefont {Burke}},
  \ and\ \bibinfo {author} {\bibfnamefont {Matthias}\ \bibnamefont
  {Ernzerhof}},\ }\bibfield  {title} {Generalized gradient approximation made
  simple,\ }\href {\doibase 10.1103/PhysRevLett.77.3865} {\bibfield  {journal}
  {\bibinfo  {journal} {Phys. Rev. Lett.}\ }\textbf {\bibinfo {volume} {77}},\
  \bibinfo {pages} {3865--3868} (\bibinfo {year} {1996})}\BibitemShut {NoStop}%
\bibitem [{\citenamefont {Tran}\ and\ \citenamefont {Blaha}(2009)}]{mbj}%
  \BibitemOpen
  \bibfield  {author} {\bibinfo {author} {\bibfnamefont {Fabien}\ \bibnamefont
  {Tran}}\ and\ \bibinfo {author} {\bibfnamefont {Peter}\ \bibnamefont
  {Blaha}},\ }\bibfield  {title} {Accurate Band Gaps of Semiconductors and
  Insulators with a Semilocal Exchange-Correlation Potential,\ }\href@noop {}
  {\bibfield  {journal} {\bibinfo  {journal} {Phys. Rev. Lett.}\ }\textbf
  {\bibinfo {volume} {102}},\ \bibinfo {pages} {226401} (\bibinfo {year}
  {2009})}\BibitemShut {NoStop}%
\bibitem [{\citenamefont {Heyd}\ \emph {et~al.}(2003)\citenamefont {Heyd},
  \citenamefont {Scuseria},\ and\ \citenamefont {Ernzerhof}}]{hse1}%
  \BibitemOpen
  \bibfield  {author} {\bibinfo {author} {\bibfnamefont {Jochen}\ \bibnamefont
  {Heyd}}, \bibinfo {author} {\bibfnamefont {Gustavo~E.}\ \bibnamefont
  {Scuseria}}, \ and\ \bibinfo {author} {\bibfnamefont {Matthias}\ \bibnamefont
  {Ernzerhof}},\ }\bibfield  {title} {Hybrid functionals based on a screened
  Coulomb potential,\ }\href@noop {} {\bibfield  {journal} {\bibinfo  {journal}
  {J. Chem. Phys.}\ }\textbf {\bibinfo {volume} {118}},\ \bibinfo {pages}
  {8207--8215} (\bibinfo {year} {2003})}\BibitemShut {NoStop}%
\bibitem [{\citenamefont {Heyd}\ \emph {et~al.}(2004)\citenamefont {Heyd},
  \citenamefont {Scuseria},\ and\ \citenamefont {Ernzerhof}}]{hse2}%
  \BibitemOpen
  \bibfield  {author} {\bibinfo {author} {\bibfnamefont {Jochen}\ \bibnamefont
  {Heyd}}, \bibinfo {author} {\bibfnamefont {Gustavo~E.}\ \bibnamefont
  {Scuseria}}, \ and\ \bibinfo {author} {\bibfnamefont {Matthias}\ \bibnamefont
  {Ernzerhof}},\ }\bibfield  {title} {Efficient hybrid density functional
  calculations in solids: Assessment of the Heyd--Scuseria--Ernzerhof screened
  Coulomb hybrid functional,\ }\href@noop {} {\bibfield  {journal} {\bibinfo
  {journal} {J. Chem. Phys.}\ }\textbf {\bibinfo {volume} {121}},\ \bibinfo
  {pages} {1187--1192} (\bibinfo {year} {2004})}\BibitemShut {NoStop}%
\bibitem [{\citenamefont {Ponc$\acute{e}$}\ \emph {et~al.}(2014)\citenamefont
  {Ponc$\acute{e}$}, \citenamefont {Antonius}, \citenamefont {Gillet},
  \citenamefont {Boulanger}, \citenamefont {Janssen}, \citenamefont {Marini},
  \citenamefont {C{\^o}t$\acute{e}$},\ and\ \citenamefont {Gonze}}]{ponceprb}%
  \BibitemOpen
  \bibfield  {author} {\bibinfo {author} {\bibfnamefont {S.}~\bibnamefont
  {Ponc$\acute{e}$}}, \bibinfo {author} {\bibfnamefont {G.}~\bibnamefont
  {Antonius}}, \bibinfo {author} {\bibfnamefont {Y.}~\bibnamefont {Gillet}},
  \bibinfo {author} {\bibfnamefont {P.}~\bibnamefont {Boulanger}}, \bibinfo
  {author} {\bibfnamefont {J.~Laflamme}\ \bibnamefont {Janssen}}, \bibinfo
  {author} {\bibfnamefont {A.}~\bibnamefont {Marini}}, \bibinfo {author}
  {\bibfnamefont {M.}~\bibnamefont {C{\^o}t$\acute{e}$}}, \ and\ \bibinfo
  {author} {\bibfnamefont {X.}~\bibnamefont {Gonze}},\ }\bibfield  {title}
  {Temperature dependence of electronic eigenenergies in the adiabatic harmonic
  approximation,\ }\href@noop {} {\bibfield  {journal} {\bibinfo  {journal}
  {Phys. Rev. B}\ }\textbf {\bibinfo {volume} {90}},\ \bibinfo {pages} {214304}
  (\bibinfo {year} {2014})}\BibitemShut {NoStop}%
\bibitem [{\citenamefont {Ponc$\acute{e}$}\ \emph {et~al.}(2015)\citenamefont
  {Ponc$\acute{e}$}, \citenamefont {Gillet}, \citenamefont {Janssen},
  \citenamefont {Marini}, \citenamefont {Verstraete},\ and\ \citenamefont
  {Gonze}}]{poncejcp}%
  \BibitemOpen
  \bibfield  {author} {\bibinfo {author} {\bibfnamefont {S.}~\bibnamefont
  {Ponc$\acute{e}$}}, \bibinfo {author} {\bibfnamefont {Y.}~\bibnamefont
  {Gillet}}, \bibinfo {author} {\bibfnamefont {J.~Laflamme}\ \bibnamefont
  {Janssen}}, \bibinfo {author} {\bibfnamefont {A.}~\bibnamefont {Marini}},
  \bibinfo {author} {\bibfnamefont {M.}~\bibnamefont {Verstraete}}, \ and\
  \bibinfo {author} {\bibfnamefont {X.}~\bibnamefont {Gonze}},\ }\bibfield
  {title} {Temperature dependence of the electronic structure of semiconductors
  and insulators,\ }\href@noop {} {\bibfield  {journal} {\bibinfo  {journal}
  {J. Chem. Phys.}\ }\textbf {\bibinfo {volume} {143}},\ \bibinfo {pages}
  {1028137} (\bibinfo {year} {2015})}\BibitemShut {NoStop}%
\bibitem [{\citenamefont {Giustino}\ \emph {et~al.}(2007)\citenamefont
  {Giustino}, \citenamefont {Cohen},\ and\ \citenamefont {Louie}}]{epw1}%
  \BibitemOpen
  \bibfield  {author} {\bibinfo {author} {\bibfnamefont {Feliciano}\
  \bibnamefont {Giustino}}, \bibinfo {author} {\bibfnamefont {Marvin~L.}\
  \bibnamefont {Cohen}}, \ and\ \bibinfo {author} {\bibfnamefont {Steven~G.}\
  \bibnamefont {Louie}},\ }\bibfield  {title} {Electron-phonon interaction
  using Wannier functions,\ }\href@noop {} {\bibfield  {journal} {\bibinfo
  {journal} {Phys. Rev. B}\ }\textbf {\bibinfo {volume} {76}},\ \bibinfo
  {pages} {165108} (\bibinfo {year} {2007})}\BibitemShut {NoStop}%
\bibitem [{\citenamefont {Noffsinger}\ \emph {et~al.}(2010)\citenamefont
  {Noffsinger}, \citenamefont {Giustino}, \citenamefont {Malone}, \citenamefont
  {Park}, \citenamefont {Louie},\ and\ \citenamefont {Cohen}}]{epw2}%
  \BibitemOpen
  \bibfield  {author} {\bibinfo {author} {\bibfnamefont {Jesse}\ \bibnamefont
  {Noffsinger}}, \bibinfo {author} {\bibfnamefont {Feliciano}\ \bibnamefont
  {Giustino}}, \bibinfo {author} {\bibfnamefont {Brad~D.}\ \bibnamefont
  {Malone}}, \bibinfo {author} {\bibfnamefont {Cheol~Hwan}\ \bibnamefont
  {Park}}, \bibinfo {author} {\bibfnamefont {Steven~G.}\ \bibnamefont {Louie}},
  \ and\ \bibinfo {author} {\bibfnamefont {Marvin~L.}\ \bibnamefont {Cohen}},\
  }\bibfield  {title} {EPW: A program for calculating the electron--phonon
  coupling using maximally localized Wannier functions,\ }\href@noop {}
  {\bibfield  {journal} {\bibinfo  {journal} {Comput. Phys. Commun.}\ }\textbf
  {\bibinfo {volume} {55}},\ \bibinfo {pages} {2140--2148} (\bibinfo {year}
  {2010})}\BibitemShut {NoStop}%
\bibitem [{\citenamefont {Ponce}\ \emph {et~al.}(2016)\citenamefont {Ponce},
  \citenamefont {Margine}, \citenamefont {Verdi},\ and\ \citenamefont
  {Giustino}}]{epw3}%
  \BibitemOpen
  \bibfield  {author} {\bibinfo {author} {\bibfnamefont {S.}~\bibnamefont
  {Ponce}}, \bibinfo {author} {\bibfnamefont {E.~R.}\ \bibnamefont {Margine}},
  \bibinfo {author} {\bibfnamefont {C.}~\bibnamefont {Verdi}}, \ and\ \bibinfo
  {author} {\bibfnamefont {F.}~\bibnamefont {Giustino}},\ }\bibfield  {title}
  {EPW: Electron--phonon coupling, transport and superconducting properties
  using maximally localized Wannier functions,\ }\href@noop {} {\bibfield
  {journal} {\bibinfo  {journal} {Comput. Phys. Commun.}\ }\textbf {\bibinfo
  {volume} {55}},\ \bibinfo {pages} {116--133} (\bibinfo {year}
  {2016})}\BibitemShut {NoStop}%
\bibitem [{\citenamefont {Giustino}(2017)}]{epwreview}%
  \BibitemOpen
  \bibfield  {author} {\bibinfo {author} {\bibfnamefont {Feliciano}\
  \bibnamefont {Giustino}},\ }\bibfield  {title} {Electron-phonon interactions
  from first principles,\ }\href@noop {} {\bibfield  {journal} {\bibinfo
  {journal} {Rev. Mod. Phys.}\ }\textbf {\bibinfo {volume} {89}},\ \bibinfo
  {pages} {015003} (\bibinfo {year} {2017})}\BibitemShut {NoStop}%
\bibitem [{\citenamefont {Marzari}\ and\ \citenamefont
  {Vanderbilt}(1997)}]{mlwfcomposite}%
  \BibitemOpen
  \bibfield  {author} {\bibinfo {author} {\bibfnamefont {Nicola}\ \bibnamefont
  {Marzari}}\ and\ \bibinfo {author} {\bibfnamefont {David}\ \bibnamefont
  {Vanderbilt}},\ }\bibfield  {title} {Maximally localized generalized Wannier
  functions for composite energy bands,\ }\href@noop {} {\bibfield  {journal}
  {\bibinfo  {journal} {Phys. Rev. B}\ }\textbf {\bibinfo {volume} {56}},\
  \bibinfo {pages} {12847--12865} (\bibinfo {year} {1997})}\BibitemShut
  {NoStop}%
\bibitem [{\citenamefont {Souza}\ \emph {et~al.}(2001)\citenamefont {Souza},
  \citenamefont {Marzari},\ and\ \citenamefont {Vanderbilt}}]{mlwfentangled}%
  \BibitemOpen
  \bibfield  {author} {\bibinfo {author} {\bibfnamefont {Ivo}\ \bibnamefont
  {Souza}}, \bibinfo {author} {\bibfnamefont {Nicola}\ \bibnamefont {Marzari}},
  \ and\ \bibinfo {author} {\bibfnamefont {David}\ \bibnamefont {Vanderbilt}},\
  }\bibfield  {title} {Maximally localized Wannier functions for entangled
  energy bands,\ }\href@noop {} {\bibfield  {journal} {\bibinfo  {journal}
  {Phys. Rev. B}\ }\textbf {\bibinfo {volume} {65}},\ \bibinfo {pages}
  {035109--1--13} (\bibinfo {year} {2001})}\BibitemShut {NoStop}%
\bibitem [{\citenamefont {Mostofi}\ \emph {et~al.}(2008)\citenamefont
  {Mostofi}, \citenamefont {Yates}, \citenamefont {Lee}, \citenamefont {Souza},
  \citenamefont {Vanderbilt},\ and\ \citenamefont {Marzari}}]{wannier90}%
  \BibitemOpen
  \bibfield  {author} {\bibinfo {author} {\bibfnamefont {Arash~A.}\
  \bibnamefont {Mostofi}}, \bibinfo {author} {\bibfnamefont {Jonathan~R.}\
  \bibnamefont {Yates}}, \bibinfo {author} {\bibfnamefont {Young-Su}\
  \bibnamefont {Lee}}, \bibinfo {author} {\bibfnamefont {Ivo}\ \bibnamefont
  {Souza}}, \bibinfo {author} {\bibfnamefont {David}\ \bibnamefont
  {Vanderbilt}}, \ and\ \bibinfo {author} {\bibfnamefont {Nicola}\ \bibnamefont
  {Marzari}},\ }\bibfield  {title} {wannier90: A tool for obtaining
  maximally-localised Wannier functions,\ }\href@noop {} {\bibfield  {journal}
  {\bibinfo  {journal} {Comput. Phys. Commun.}\ }\textbf {\bibinfo {volume}
  {178}},\ \bibinfo {pages} {685--699} (\bibinfo {year} {2008})}\BibitemShut
  {NoStop}%
\bibitem [{\citenamefont {Verdi}\ and\ \citenamefont
  {Giustino}(2015)}]{epwpolar}%
  \BibitemOpen
  \bibfield  {author} {\bibinfo {author} {\bibfnamefont {C.}~\bibnamefont
  {Verdi}}\ and\ \bibinfo {author} {\bibfnamefont {F.}~\bibnamefont
  {Giustino}},\ }\bibfield  {title} {Fr{\"o}hlich Electron-Phonon Vertex from
  First Principles,\ }\href@noop {} {\bibfield  {journal} {\bibinfo  {journal}
  {Phys. Rev. Lett.}\ }\textbf {\bibinfo {volume} {115}},\ \bibinfo {pages}
  {176401} (\bibinfo {year} {2015})}\BibitemShut {NoStop}%
\bibitem [{sup()}]{supplementary}%
  \BibitemOpen
  \href@noop {} {\bibinfo  {journal} {Supplementary materials to the main text
  for computational details on electron-phonon scattering, phase stability and
  defects}\ }\BibitemShut {NoStop}%
\bibitem [{\citenamefont {Madsen}\ and\ \citenamefont
  {Singh}(2006)}]{boltztrap}%
  \BibitemOpen
\bibfield  {journal} {  }\bibfield  {author} {\bibinfo {author} {\bibfnamefont
  {Georg~K.H.}\ \bibnamefont {Madsen}}\ and\ \bibinfo {author} {\bibfnamefont
  {David~J.}\ \bibnamefont {Singh}},\ }\bibfield  {title} {BoltzTraP. A code
  for calculating band-structure dependent quantities,\ }\href@noop {}
  {\bibfield  {journal} {\bibinfo  {journal} {Comput. Phys. Commun.}\ }\textbf
  {\bibinfo {volume} {175}},\ \bibinfo {pages} {67--71} (\bibinfo {year}
  {2006})}\BibitemShut {NoStop}%
\bibitem [{\citenamefont {Song}\ \emph {et~al.}(2017)\citenamefont {Song},
  \citenamefont {Liu}, \citenamefont {Zhou}, \citenamefont {Ding},\ and\
  \citenamefont {Chen}}]{pbteepw1}%
  \BibitemOpen
  \bibfield  {author} {\bibinfo {author} {\bibfnamefont {Qichen}\ \bibnamefont
  {Song}}, \bibinfo {author} {\bibfnamefont {Te-Huan}\ \bibnamefont {Liu}},
  \bibinfo {author} {\bibfnamefont {Jiawei}\ \bibnamefont {Zhou}}, \bibinfo
  {author} {\bibfnamefont {Zhiwei}\ \bibnamefont {Ding}}, \ and\ \bibinfo
  {author} {\bibfnamefont {Gang}\ \bibnamefont {Chen}},\ }\bibfield  {title}
  {\textit{Ab initio} study of electron mean free paths and thermoelectric
  properties of lead telluride,\ }\href@noop {} {\bibfield  {journal} {\bibinfo
   {journal} {Mater. Today Phys.}\ }\textbf {\bibinfo {volume} {2}},\ \bibinfo
  {pages} {69--77} (\bibinfo {year} {2017})}\BibitemShut {NoStop}%
\bibitem [{\citenamefont {Cao}\ \emph {et~al.}(2018)\citenamefont {Cao},
  \citenamefont {Querales-Flores}, \citenamefont {Murphy}, \citenamefont
  {Fahy},\ and\ \citenamefont {Savi{\'c}}}]{pbteepw2}%
  \BibitemOpen
  \bibfield  {author} {\bibinfo {author} {\bibfnamefont {Jiang}\ \bibnamefont
  {Cao}}, \bibinfo {author} {\bibfnamefont {Jos{\'e}~D.}\ \bibnamefont
  {Querales-Flores}}, \bibinfo {author} {\bibfnamefont {Aoife~R.}\ \bibnamefont
  {Murphy}}, \bibinfo {author} {\bibfnamefont {Stephen}\ \bibnamefont {Fahy}},
  \ and\ \bibinfo {author} {\bibfnamefont {Ivana}\ \bibnamefont {Savi{\'c}}},\
  }\bibfield  {title} {Dominant electron-phonon scattering mechanisms in
  $n$-type PbTe from first principles,\ }\href@noop {} {\bibfield  {journal}
  {\bibinfo  {journal} {Phys. Rev. B}\ }\textbf {\bibinfo {volume} {98}},\
  \bibinfo {pages} {205202} (\bibinfo {year} {2018})}\BibitemShut {NoStop}%
\bibitem [{\citenamefont {Zhou}\ \emph {et~al.}(2018)\citenamefont {Zhou},
  \citenamefont {Zhu}, \citenamefont {Liu}, \citenamefont {Song}, \citenamefont
  {He}, \citenamefont {Mao}, \citenamefont {Liu}, \citenamefont {Ren},
  \citenamefont {Liao}, \citenamefont {Singh},\ and\ \citenamefont
  {Chen}}]{nbfesbnatcomm}%
  \BibitemOpen
  \bibfield  {author} {\bibinfo {author} {\bibfnamefont {J.}~\bibnamefont
  {Zhou}}, \bibinfo {author} {\bibfnamefont {H.}~\bibnamefont {Zhu}}, \bibinfo
  {author} {\bibfnamefont {T.}~\bibnamefont {Liu}}, \bibinfo {author}
  {\bibfnamefont {Q.}~\bibnamefont {Song}}, \bibinfo {author} {\bibfnamefont
  {R.}~\bibnamefont {He}}, \bibinfo {author} {\bibfnamefont {J.}~\bibnamefont
  {Mao}}, \bibinfo {author} {\bibfnamefont {Z.}~\bibnamefont {Liu}}, \bibinfo
  {author} {\bibfnamefont {W.}~\bibnamefont {Ren}}, \bibinfo {author}
  {\bibfnamefont {B.}~\bibnamefont {Liao}}, \bibinfo {author} {\bibfnamefont
  {D.~J.}\ \bibnamefont {Singh}}, \ and\ \bibinfo {author} {\bibfnamefont
  {G.}~\bibnamefont {Chen}},\ }\bibfield  {title} {Large thermoelectric power
  factor from crystal symmetry-protected non-bonding orbital in half-Heuslers
  Nb$_{1-x}$Ti$_{x}$FeSb,\ }\href@noop {} {\bibfield  {journal} {\bibinfo
  {journal} {Nat. Commun.}\ }\textbf {\bibinfo {volume} {9}},\ \bibinfo {pages}
  {1--9} (\bibinfo {year} {2018})}\BibitemShut {NoStop}%
\bibitem [{\citenamefont {Xia}\ \emph {et~al.}(2019)\citenamefont {Xia},
  \citenamefont {Park}, \citenamefont {Zhou},\ and\ \citenamefont
  {Ozoli\c{n}\v{s}}}]{cosiyi}%
  \BibitemOpen
  \bibfield  {author} {\bibinfo {author} {\bibfnamefont {Yi}~\bibnamefont
  {Xia}}, \bibinfo {author} {\bibfnamefont {Junsoo}\ \bibnamefont {Park}},
  \bibinfo {author} {\bibfnamefont {Fei}\ \bibnamefont {Zhou}}, \ and\ \bibinfo
  {author} {\bibfnamefont {Vidivuds}\ \bibnamefont {Ozoli\c{n}\v{s}}},\
  }\bibfield  {title} {High Thermoelectric Power Factor in Intermetallic CoSi
  Arising from Energy Filtering of Electrons by Phonon Scattering,\ }\href@noop
  {} {\bibfield  {journal} {\bibinfo  {journal} {Phys. Rev. Appl.}\ }\textbf
  {\bibinfo {volume} {11}},\ \bibinfo {pages} {024017} (\bibinfo {year}
  {2019})}\BibitemShut {NoStop}%
\bibitem [{\citenamefont {Kresse}\ and\ \citenamefont {Hafner}(1993)}]{vasp1}%
  \BibitemOpen
  \bibfield  {author} {\bibinfo {author} {\bibfnamefont {G.}~\bibnamefont
  {Kresse}}\ and\ \bibinfo {author} {\bibfnamefont {J.}~\bibnamefont
  {Hafner}},\ }\bibfield  {title} {\textit{Ab initio} molecular dynamics for
  liquid metals,\ }\href {\doibase 10.1103/PhysRevB.47.558} {\bibfield
  {journal} {\bibinfo  {journal} {Phys. Rev. B}\ }\textbf {\bibinfo {volume}
  {47}},\ \bibinfo {pages} {558--561} (\bibinfo {year} {1993})}\BibitemShut
  {NoStop}%
\bibitem [{\citenamefont {Kresse}\ and\ \citenamefont {Hafner}(1994)}]{vasp2}%
  \BibitemOpen
  \bibfield  {author} {\bibinfo {author} {\bibfnamefont {G.}~\bibnamefont
  {Kresse}}\ and\ \bibinfo {author} {\bibfnamefont {J.}~\bibnamefont
  {Hafner}},\ }\bibfield  {title} {\textit{Ab initio} molecular-dynamics
  simulation of the liquid-metal\char21{}amorphous-semiconductor transition in
  germanium,\ }\href@noop {} {\bibfield  {journal} {\bibinfo  {journal} {Phys.
  Rev. B}\ }\textbf {\bibinfo {volume} {49}},\ \bibinfo {pages} {14251--14269}
  (\bibinfo {year} {1994})}\BibitemShut {NoStop}%
\bibitem [{\citenamefont {Kresse}\ and\ \citenamefont
  {Furthm{\"u}ller}(1996)}]{vasp3}%
  \BibitemOpen
  \bibfield  {author} {\bibinfo {author} {\bibfnamefont {G.}~\bibnamefont
  {Kresse}}\ and\ \bibinfo {author} {\bibfnamefont {J.}~\bibnamefont
  {Furthm{\"u}ller}},\ }\bibfield  {title} {Efficiency of ab-initio total
  energy calculations for metals and semiconductors using a plane-wave basis
  set,\ }\href@noop {} {\bibfield  {journal} {\bibinfo  {journal} {Comput.
  Mater. Sci.}\ }\textbf {\bibinfo {volume} {6}},\ \bibinfo {pages} {15--50}
  (\bibinfo {year} {1996})}\BibitemShut {NoStop}%
\bibitem [{\citenamefont {Kresse}\ and\ \citenamefont
  {Furthm\"uller}(1996)}]{vasp4}%
  \BibitemOpen
  \bibfield  {author} {\bibinfo {author} {\bibfnamefont {G.}~\bibnamefont
  {Kresse}}\ and\ \bibinfo {author} {\bibfnamefont {J.}~\bibnamefont
  {Furthm\"uller}},\ }\bibfield  {title} {Efficient iterative schemes for
  \textit{ab initio} total-energy calculations using a plane-wave basis set,\
  }\href@noop {} {\bibfield  {journal} {\bibinfo  {journal} {Phys. Rev. B}\
  }\textbf {\bibinfo {volume} {54}},\ \bibinfo {pages} {11169--11186} (\bibinfo
  {year} {1996})}\BibitemShut {NoStop}%
\bibitem [{\citenamefont {Bl\"ochl}(1994)}]{paw}%
  \BibitemOpen
  \bibfield  {author} {\bibinfo {author} {\bibfnamefont {P.~E.}\ \bibnamefont
  {Bl\"ochl}},\ }\bibfield  {title} {Projector augmented-wave method,\ }\href
  {\doibase 10.1103/PhysRevB.50.17953} {\bibfield  {journal} {\bibinfo
  {journal} {Phys. Rev. B}\ }\textbf {\bibinfo {volume} {50}},\ \bibinfo
  {pages} {17953--17979} (\bibinfo {year} {1994})}\BibitemShut {NoStop}%
\bibitem [{\citenamefont {Jain}\ \emph {et~al.}(2013)\citenamefont {Jain},
  \citenamefont {Ong}, \citenamefont {Hautier}, \citenamefont {Chen},
  \citenamefont {Richards}, \citenamefont {Dacek}, \citenamefont {Cholia},
  \citenamefont {Gunter}, \citenamefont {Skinner}, \citenamefont {Ceder},\ and\
  \citenamefont {Perssson}}]{materialsproject}%
  \BibitemOpen
  \bibfield  {author} {\bibinfo {author} {\bibfnamefont {A.}~\bibnamefont
  {Jain}}, \bibinfo {author} {\bibfnamefont {S.~P.}\ \bibnamefont {Ong}},
  \bibinfo {author} {\bibfnamefont {G.}~\bibnamefont {Hautier}}, \bibinfo
  {author} {\bibfnamefont {W.}~\bibnamefont {Chen}}, \bibinfo {author}
  {\bibfnamefont {W.~D.}\ \bibnamefont {Richards}}, \bibinfo {author}
  {\bibfnamefont {S.}~\bibnamefont {Dacek}}, \bibinfo {author} {\bibfnamefont
  {S.}~\bibnamefont {Cholia}}, \bibinfo {author} {\bibfnamefont
  {D.}~\bibnamefont {Gunter}}, \bibinfo {author} {\bibfnamefont
  {D.}~\bibnamefont {Skinner}}, \bibinfo {author} {\bibfnamefont
  {G.}~\bibnamefont {Ceder}}, \ and\ \bibinfo {author} {\bibfnamefont {K.~A.}\
  \bibnamefont {Perssson}},\ }\bibfield  {title} {Commentary: The Materials
  Project: A materials genome approach to accelerating materials innovation,\
  }\href@noop {} {\bibfield  {journal} {\bibinfo  {journal} {APL Mater.}\
  }\textbf {\bibinfo {volume} {1}},\ \bibinfo {pages} {1002} (\bibinfo {year}
  {2013})}\BibitemShut {NoStop}%
\bibitem [{\citenamefont {Bergerhoff}\ and\ \citenamefont
  {Brown}(1987)}]{icsd1}%
  \BibitemOpen
  \bibfield  {author} {\bibinfo {author} {\bibfnamefont {G.}~\bibnamefont
  {Bergerhoff}}\ and\ \bibinfo {author} {\bibfnamefont {I.~D.}\ \bibnamefont
  {Brown}},\ }\bibfield  {title} {Inorganic crystal structure database,\
  }\href@noop {} {\bibfield  {journal} {\bibinfo  {journal} {Crystallographic
  Databases}\ ,\ \bibinfo {pages} {77--95}} (\bibinfo {year}
  {1987})}\BibitemShut {NoStop}%
\bibitem [{\citenamefont {Belsky}\ \emph {et~al.}(2002)\citenamefont {Belsky},
  \citenamefont {Hellenbrandt}, \citenamefont {Karen},\ and\ \citenamefont
  {Luksch}}]{icsd2}%
  \BibitemOpen
  \bibfield  {author} {\bibinfo {author} {\bibfnamefont {Alex}\ \bibnamefont
  {Belsky}}, \bibinfo {author} {\bibfnamefont {Mariette}\ \bibnamefont
  {Hellenbrandt}}, \bibinfo {author} {\bibfnamefont {Vicky~Lynn}\ \bibnamefont
  {Karen}}, \ and\ \bibinfo {author} {\bibfnamefont {Peter}\ \bibnamefont
  {Luksch}},\ }\bibfield  {title} {New developments in the Inorganic Crystal
  Structure Database (ICSD): accessibility in support of materials research and
  design,\ }\href@noop {} {\bibfield  {journal} {\bibinfo  {journal} {Acta
  Crystallogr. B}\ }\textbf {\bibinfo {volume} {58}},\ \bibinfo {pages}
  {364--369} (\bibinfo {year} {2002})}\BibitemShut {NoStop}%
\bibitem [{\citenamefont {Hellenbrandt}(2004)}]{icsd3}%
  \BibitemOpen
  \bibfield  {author} {\bibinfo {author} {\bibfnamefont {Mariette}\
  \bibnamefont {Hellenbrandt}},\ }\bibfield  {title} {The Inorganic Crystal
  Structure Databae (ICSD) -- Present and Future,\ }\href@noop {} {\bibfield
  {journal} {\bibinfo  {journal} {Crystallogr. Rev.}\ }\textbf {\bibinfo
  {volume} {10}},\ \bibinfo {pages} {17--22} (\bibinfo {year}
  {2004})}\BibitemShut {NoStop}%
\bibitem [{\citenamefont {Methfessel}\ and\ \citenamefont
  {Paxton}(1989)}]{mpmethod}%
  \BibitemOpen
  \bibfield  {author} {\bibinfo {author} {\bibfnamefont {M.}~\bibnamefont
  {Methfessel}}\ and\ \bibinfo {author} {\bibfnamefont {A.T.}\ \bibnamefont
  {Paxton}},\ }\bibfield  {title} {High-precision sampling for Brillouin-zone
  integration in metals,\ }\href@noop {} {\bibfield  {journal} {\bibinfo
  {journal} {Phys. Rev. B}\ }\textbf {\bibinfo {volume} {40}},\ \bibinfo
  {pages} {3616--3621} (\bibinfo {year} {1989})}\BibitemShut {NoStop}%
\bibitem [{\citenamefont {Makov}\ and\ \citenamefont
  {Payne}(1995)}]{makovpayne}%
  \BibitemOpen
  \bibfield  {author} {\bibinfo {author} {\bibfnamefont {G.}~\bibnamefont
  {Makov}}\ and\ \bibinfo {author} {\bibfnamefont {M.~C.}\ \bibnamefont
  {Payne}},\ }\bibfield  {title} {Periodic boundary conditions in \textit{ab
  initio} calculations,\ }\href@noop {} {\bibfield  {journal} {\bibinfo
  {journal} {Phys. Rev. B}\ }\textbf {\bibinfo {volume} {51}},\ \bibinfo
  {pages} {074014} (\bibinfo {year} {1995})}\BibitemShut {NoStop}%
\bibitem [{\citenamefont {Freysoldt}\ \emph {et~al.}(2009)\citenamefont
  {Freysoldt}, \citenamefont {Neugebauer},\ and\ \citenamefont
  {de~Walle}}]{supercellsizeprl}%
  \BibitemOpen
  \bibfield  {author} {\bibinfo {author} {\bibfnamefont {Christoph}\
  \bibnamefont {Freysoldt}}, \bibinfo {author} {\bibfnamefont
  {J$\ddot{\text{o}}$rg}\ \bibnamefont {Neugebauer}}, \ and\ \bibinfo {author}
  {\bibfnamefont {Chris G.~Van}\ \bibnamefont {de~Walle}},\ }\bibfield  {title}
  {Fully \textit{Ab Initio} Finite-Size Corrections for Charged-Defect
  Supercell Calculations,\ }\href@noop {} {\bibfield  {journal} {\bibinfo
  {journal} {Phys. Rev. B}\ }\textbf {\bibinfo {volume} {102}},\ \bibinfo
  {pages} {016402} (\bibinfo {year} {2009})}\BibitemShut {NoStop}%
\bibitem [{\citenamefont {Freysoldt}\ \emph {et~al.}(2011)\citenamefont
  {Freysoldt}, \citenamefont {Neugebauer},\ and\ \citenamefont
  {de~Walle}}]{supercellsizepss}%
  \BibitemOpen
  \bibfield  {author} {\bibinfo {author} {\bibfnamefont {Christoph}\
  \bibnamefont {Freysoldt}}, \bibinfo {author} {\bibfnamefont
  {J$\ddot{\text{o}}$rg}\ \bibnamefont {Neugebauer}}, \ and\ \bibinfo {author}
  {\bibfnamefont {Chris G.~Van}\ \bibnamefont {de~Walle}},\ }\bibfield  {title}
  {Electrostatic interactions between charged defects in supercells,\
  }\href@noop {} {\bibfield  {journal} {\bibinfo  {journal} {Phys. Stat.
  Solid.}\ }\textbf {\bibinfo {volume} {248}},\ \bibinfo {pages} {1067}
  (\bibinfo {year} {2011})}\BibitemShut {NoStop}%
\bibitem [{\citenamefont {Lany}\ and\ \citenamefont
  {Zunger}(2008)}]{lanyzunger}%
  \BibitemOpen
  \bibfield  {author} {\bibinfo {author} {\bibfnamefont {Stephan}\ \bibnamefont
  {Lany}}\ and\ \bibinfo {author} {\bibfnamefont {Alex}\ \bibnamefont
  {Zunger}},\ }\bibfield  {title} {Assessment of correction methods for the
  band-gap problem and for finite-size effects in supercell defect
  calculations: Case studies for ZnO and GaAs,\ }\href@noop {} {\bibfield
  {journal} {\bibinfo  {journal} {Phys. Rev. B}\ }\textbf {\bibinfo {volume}
  {78}},\ \bibinfo {pages} {235104} (\bibinfo {year} {2008})}\BibitemShut
  {NoStop}%
\bibitem [{\citenamefont {Alkauskas}\ \emph {et~al.}(2008)\citenamefont
  {Alkauskas}, \citenamefont {Broqvist},\ and\ \citenamefont
  {Pasquarello}}]{defecthse1}%
  \BibitemOpen
  \bibfield  {author} {\bibinfo {author} {\bibfnamefont {Audrius}\ \bibnamefont
  {Alkauskas}}, \bibinfo {author} {\bibfnamefont {Peter}\ \bibnamefont
  {Broqvist}}, \ and\ \bibinfo {author} {\bibfnamefont {Alfredo}\ \bibnamefont
  {Pasquarello}},\ }\bibfield  {title} {Defect Energy Levels in Density
  Functional Calculations: Alignment and Band Gap Problem,\ }\href@noop {}
  {\bibfield  {journal} {\bibinfo  {journal} {Phys. Rev. B}\ }\textbf {\bibinfo
  {volume} {101}},\ \bibinfo {pages} {046404} (\bibinfo {year}
  {2008})}\BibitemShut {NoStop}%
\bibitem [{\citenamefont {Komsa}\ \emph {et~al.}(2010)\citenamefont {Komsa},
  \citenamefont {Broqvist},\ and\ \citenamefont {Pasquarello}}]{defecthse2}%
  \BibitemOpen
  \bibfield  {author} {\bibinfo {author} {\bibfnamefont {Hannu-Pekka}\
  \bibnamefont {Komsa}}, \bibinfo {author} {\bibfnamefont {Peter}\ \bibnamefont
  {Broqvist}}, \ and\ \bibinfo {author} {\bibfnamefont {Alfredo}\ \bibnamefont
  {Pasquarello}},\ }\bibfield  {title} {Alignment of defect levels and band
  edges through hybrid functionals: Effect of screening in the exchange term,\
  }\href@noop {} {\bibfield  {journal} {\bibinfo  {journal} {Phys. Rev. B}\
  }\textbf {\bibinfo {volume} {81}},\ \bibinfo {pages} {205118} (\bibinfo
  {year} {2010})}\BibitemShut {NoStop}%
\bibitem [{\citenamefont {Komsa}\ and\ \citenamefont
  {Pasquarello}(2011)}]{defecthse3}%
  \BibitemOpen
  \bibfield  {author} {\bibinfo {author} {\bibfnamefont {Hannu-Pekka}\
  \bibnamefont {Komsa}}\ and\ \bibinfo {author} {\bibfnamefont {Alfredo}\
  \bibnamefont {Pasquarello}},\ }\bibfield  {title} {Assessing the accuracy of
  hybrid functionals in the determination of defect levels: Application to the
  As antisite in GaAs,\ }\href@noop {} {\bibfield  {journal} {\bibinfo
  {journal} {Phys. Rev. B}\ }\textbf {\bibinfo {volume} {84}},\ \bibinfo
  {pages} {075207} (\bibinfo {year} {2011})}\BibitemShut {NoStop}%
\bibitem [{\citenamefont {Alkauskas}\ and\ \citenamefont
  {Pasquarello}(2011)}]{defecthse4}%
  \BibitemOpen
  \bibfield  {author} {\bibinfo {author} {\bibfnamefont {Audrius}\ \bibnamefont
  {Alkauskas}}\ and\ \bibinfo {author} {\bibfnamefont {Alfredo}\ \bibnamefont
  {Pasquarello}},\ }\bibfield  {title} {Band-edge problem in the theoretical
  determination of defect energy levels: The O vacancy in ZnO as a benchmark
  case,\ }\href@noop {} {\bibfield  {journal} {\bibinfo  {journal} {Phys. Rev.
  B}\ }\textbf {\bibinfo {volume} {84}},\ \bibinfo {pages} {125206} (\bibinfo
  {year} {2011})}\BibitemShut {NoStop}%
\bibitem [{\citenamefont {Du}(2015)}]{defecthse5}%
  \BibitemOpen
  \bibfield  {author} {\bibinfo {author} {\bibfnamefont {Mao-Hua}\ \bibnamefont
  {Du}},\ }\bibfield  {title} {Density Functional Calculations of Native
  Defects in CH$_{3}$NH$_{3}$PbI$_{3}$: Effects of Spin-Orbit Coupling and
  Self-Interaction Error,\ }\href@noop {} {\bibfield  {journal} {\bibinfo
  {journal} {J. Phys. Chem. Lett.}\ }\textbf {\bibinfo {volume} {6}},\ \bibinfo
  {pages} {1461--1466} (\bibinfo {year} {2015})}\BibitemShut {NoStop}%
\bibitem [{\citenamefont {He}\ \emph {et~al.}(2016{\natexlab{b}})\citenamefont
  {He}, \citenamefont {Kraemer}, \citenamefont {Mao}, \citenamefont {Zeng},
  \citenamefont {Jie}, \citenamefont {Lan}, \citenamefont {Lie}, \citenamefont
  {Shuai}, \citenamefont {Kim}, \citenamefont {Liu}, \citenamefont {Broido},
  \citenamefont {Chu}, \citenamefont {Chen},\ and\ \citenamefont
  {Ren}}]{nbfesbpnas}%
  \BibitemOpen
  \bibfield  {author} {\bibinfo {author} {\bibfnamefont {Ran}\ \bibnamefont
  {He}}, \bibinfo {author} {\bibfnamefont {Daniel}\ \bibnamefont {Kraemer}},
  \bibinfo {author} {\bibfnamefont {Jun}\ \bibnamefont {Mao}}, \bibinfo
  {author} {\bibfnamefont {Lingping}\ \bibnamefont {Zeng}}, \bibinfo {author}
  {\bibfnamefont {Qing}\ \bibnamefont {Jie}}, \bibinfo {author} {\bibfnamefont
  {Yucheng}\ \bibnamefont {Lan}}, \bibinfo {author} {\bibfnamefont {Chunhua}\
  \bibnamefont {Lie}}, \bibinfo {author} {\bibfnamefont {Jing}\ \bibnamefont
  {Shuai}}, \bibinfo {author} {\bibfnamefont {Hee~Seok}\ \bibnamefont {Kim}},
  \bibinfo {author} {\bibfnamefont {Yuan}\ \bibnamefont {Liu}}, \bibinfo
  {author} {\bibfnamefont {David}\ \bibnamefont {Broido}}, \bibinfo {author}
  {\bibfnamefont {Ching-Wu}\ \bibnamefont {Chu}}, \bibinfo {author}
  {\bibfnamefont {Gang}\ \bibnamefont {Chen}}, \ and\ \bibinfo {author}
  {\bibfnamefont {Zhifeng}\ \bibnamefont {Ren}},\ }\bibfield  {title}
  {Achieving high power factor and output power density in $p$-type
  half-Heuslers Nb$_{1-x}$Ti$_{x}$FeSb,\ }\href@noop {} {\bibfield  {journal}
  {\bibinfo  {journal} {Proc. Natl. Acad. Sci.}\ }\textbf {\bibinfo {volume}
  {113}},\ \bibinfo {pages} {13576--13581} (\bibinfo {year}
  {2016}{\natexlab{b}})}\BibitemShut {NoStop}%
\bibitem [{\citenamefont {Sootsman}\ \emph {et~al.}(2008)\citenamefont
  {Sootsman}, \citenamefont {Kong}, \citenamefont {Uher}, \citenamefont
  {Angelo}, \citenamefont {Wu}, \citenamefont {Hogan}, \citenamefont
  {Caillat},\ and\ \citenamefont {Kanatzidis}}]{pbtenanostructuring}%
  \BibitemOpen
  \bibfield  {author} {\bibinfo {author} {\bibfnamefont {Joseph~R.}\
  \bibnamefont {Sootsman}}, \bibinfo {author} {\bibfnamefont {Huijun}\
  \bibnamefont {Kong}}, \bibinfo {author} {\bibfnamefont {Ctirad}\ \bibnamefont
  {Uher}}, \bibinfo {author} {\bibfnamefont {Jonathan James~D}\ \bibnamefont
  {Angelo}}, \bibinfo {author} {\bibfnamefont {Chun-I}\ \bibnamefont {Wu}},
  \bibinfo {author} {\bibfnamefont {Timothy~P.}\ \bibnamefont {Hogan}},
  \bibinfo {author} {\bibfnamefont {Thierry}\ \bibnamefont {Caillat}}, \ and\
  \bibinfo {author} {\bibfnamefont {Mercouri~G.}\ \bibnamefont {Kanatzidis}},\
  }\bibfield  {title} {Large Enhancements in the Thermoelectric Power Factor of
  Bulk PbTe at High Temperature by Synergistic Nanostructuring,\ }\href@noop {}
  {\bibfield  {journal} {\bibinfo  {journal} {Angew. Chem.}\ }\textbf {\bibinfo
  {volume} {47}},\ \bibinfo {pages} {8618--8622} (\bibinfo {year}
  {2008})}\BibitemShut {NoStop}%
\bibitem [{\citenamefont {Pei}\ \emph {et~al.}(2011)\citenamefont {Pei},
  \citenamefont {Xiaoya}, \citenamefont {Aaron}, \citenamefont {H.~Wang},\ and\
  \citenamefont {Snyder}}]{pbtebandconvergence}%
  \BibitemOpen
  \bibfield  {author} {\bibinfo {author} {\bibfnamefont {Y.}~\bibnamefont
  {Pei}}, \bibinfo {author} {\bibfnamefont {S.}~\bibnamefont {Xiaoya}},
  \bibinfo {author} {\bibfnamefont {L.}~\bibnamefont {Aaron}}, \bibinfo
  {author} {\bibfnamefont {L.~Chen}\ \bibnamefont {H.~Wang}}, \ and\ \bibinfo
  {author} {\bibfnamefont {G.J.}\ \bibnamefont {Snyder}},\ }\bibfield  {title}
  {Convergence of electronic bands for high performance bulk thermoelectrics,\
  }\href@noop {} {\bibfield  {journal} {\bibinfo  {journal} {Nature}\ }\textbf
  {\bibinfo {volume} {473}},\ \bibinfo {pages} {66--69} (\bibinfo {year}
  {2011})}\BibitemShut {NoStop}%
\bibitem [{\citenamefont {Pei}\ \emph {et~al.}(2012{\natexlab{b}})\citenamefont
  {Pei}, \citenamefont {Wang},\ and\ \citenamefont
  {Snyder}}]{bandconvergencereview}%
  \BibitemOpen
  \bibfield  {author} {\bibinfo {author} {\bibfnamefont {Y.}~\bibnamefont
  {Pei}}, \bibinfo {author} {\bibfnamefont {H.}~\bibnamefont {Wang}}, \ and\
  \bibinfo {author} {\bibfnamefont {G.J.}\ \bibnamefont {Snyder}},\ }\bibfield
  {title} {Band engineering of thermoelectric materials,\ }\href@noop {}
  {\bibfield  {journal} {\bibinfo  {journal} {Adv. Mater.}\ }\textbf {\bibinfo
  {volume} {24}},\ \bibinfo {pages} {6125--6135} (\bibinfo {year}
  {2012}{\natexlab{b}})}\BibitemShut {NoStop}%
\bibitem [{\citenamefont {Biswas}\ \emph {et~al.}(2012)\citenamefont {Biswas},
  \citenamefont {He}, \citenamefont {Blum}, \citenamefont {Wu}, \citenamefont
  {Hogan}, \citenamefont {abd Vinayak P.~Dravid},\ and\ \citenamefont
  {Kanatzidis}}]{pbsrte}%
  \BibitemOpen
  \bibfield  {author} {\bibinfo {author} {\bibfnamefont {Kanishka}\
  \bibnamefont {Biswas}}, \bibinfo {author} {\bibfnamefont {Jiaqing}\
  \bibnamefont {He}}, \bibinfo {author} {\bibfnamefont {Ivan~D.}\ \bibnamefont
  {Blum}}, \bibinfo {author} {\bibfnamefont {Chun-I}\ \bibnamefont {Wu}},
  \bibinfo {author} {\bibfnamefont {Timothy~P.}\ \bibnamefont {Hogan}},
  \bibinfo {author} {\bibfnamefont {David N.~Seidman}\ \bibnamefont {abd
  Vinayak P.~Dravid}}, \ and\ \bibinfo {author} {\bibfnamefont {Mercouri~G.}\
  \bibnamefont {Kanatzidis}},\ }\bibfield  {title} {High-performance bulk
  thermoelectrics with all-scale hierarchical architectures,\ }\href@noop {}
  {\bibfield  {journal} {\bibinfo  {journal} {Nature}\ }\textbf {\bibinfo
  {volume} {489}},\ \bibinfo {pages} {414--418} (\bibinfo {year}
  {2012})}\BibitemShut {NoStop}%
\bibitem [{\citenamefont {Smith}\ and\ \citenamefont
  {Wolfe}(1962)}]{bisbsmithwolfe}%
  \BibitemOpen
  \bibfield  {author} {\bibinfo {author} {\bibfnamefont {G.~E.}\ \bibnamefont
  {Smith}}\ and\ \bibinfo {author} {\bibfnamefont {R.}~\bibnamefont {Wolfe}},\
  }\bibfield  {title} {Thermoelectric Properties of Bismuth- Antimony Alloys,\
  }\href@noop {} {\bibfield  {journal} {\bibinfo  {journal} {J. Appl. Phys.}\
  }\textbf {\bibinfo {volume} {33}},\ \bibinfo {pages} {841--846} (\bibinfo
  {year} {1962})}\BibitemShut {NoStop}%
\bibitem [{\citenamefont {Nolas}\ \emph {et~al.}(2001)\citenamefont {Nolas},
  \citenamefont {Sharp},\ and\ \citenamefont {Goldsmid}}]{thermoelectrics}%
  \BibitemOpen
  \bibfield  {author} {\bibinfo {author} {\bibfnamefont {G.~S.}\ \bibnamefont
  {Nolas}}, \bibinfo {author} {\bibfnamefont {J.}~\bibnamefont {Sharp}}, \ and\
  \bibinfo {author} {\bibfnamefont {H.~J.}\ \bibnamefont {Goldsmid}},\
  }\href@noop {} {\emph {\bibinfo {title} {Thermoelectrics}}}\ (\bibinfo
  {publisher} {Springer},\ \bibinfo {year} {2001})\BibitemShut {NoStop}%
\bibitem [{\citenamefont {Zhou}\ \emph {et~al.}(2014)\citenamefont {Zhou},
  \citenamefont {Nielson}, \citenamefont {Xia},\ and\ \citenamefont
  {Ozoli\c{n}\v{s}}}]{csld}%
  \BibitemOpen
  \bibfield  {author} {\bibinfo {author} {\bibfnamefont {F.}~\bibnamefont
  {Zhou}}, \bibinfo {author} {\bibfnamefont {W.}~\bibnamefont {Nielson}},
  \bibinfo {author} {\bibfnamefont {Y.}~\bibnamefont {Xia}}, \ and\ \bibinfo
  {author} {\bibfnamefont {V.}~\bibnamefont {Ozoli\c{n}\v{s}}},\ }\bibfield
  {title} {Lattice Anharmonicity and Thermal Conductivity from Compressive
  Sensing of First-Principles Calculations,\ }\href@noop {} {\bibfield
  {journal} {\bibinfo  {journal} {Phys. Rev. Lett.}\ }\textbf {\bibinfo
  {volume} {113}},\ \bibinfo {pages} {185501} (\bibinfo {year}
  {2014})}\BibitemShut {NoStop}%
\bibitem [{\citenamefont {Li}\ \emph {et~al.}(2014)\citenamefont {Li},
  \citenamefont {Carrete}, \citenamefont {Katcho},\ and\ \citenamefont
  {Mingo}}]{shengbte}%
  \BibitemOpen
  \bibfield  {author} {\bibinfo {author} {\bibfnamefont {Wu}~\bibnamefont
  {Li}}, \bibinfo {author} {\bibfnamefont {Jes\'us}\ \bibnamefont {Carrete}},
  \bibinfo {author} {\bibfnamefont {Nebil~A.}\ \bibnamefont {Katcho}}, \ and\
  \bibinfo {author} {\bibfnamefont {Natalio}\ \bibnamefont {Mingo}},\
  }\bibfield  {title} {ShengBTE: A solver of the Boltzmann transport equation
  for phonons,\ }\href {\doibase 10.1016/j.cpc.2014.02.015} {\bibfield
  {journal} {\bibinfo  {journal} {Comput. Phys. Commun.}\ }\textbf {\bibinfo
  {volume} {185}},\ \bibinfo {pages} {1747--1758} (\bibinfo {year}
  {2014})}\BibitemShut {NoStop}%
\bibitem [{\citenamefont {Zunger}(2019)}]{fantasymaterials}%
  \BibitemOpen
  \bibfield  {author} {\bibinfo {author} {\bibfnamefont {Alex}\ \bibnamefont
  {Zunger}},\ }\bibfield  {title} {Beware of plausible predictions of fantasy
  materials,\ }\href@noop {} {\bibfield  {journal} {\bibinfo  {journal}
  {Nature}\ }\textbf {\bibinfo {volume} {566}},\ \bibinfo {pages} {447--449}
  (\bibinfo {year} {2019})}\BibitemShut {NoStop}%
\bibitem [{\citenamefont {V}\ \emph {et~al.}(1990)\citenamefont {V},
  \citenamefont {Burch}, \citenamefont {Rag},\ and\ \citenamefont
  {Budnick}}]{rmxcompounds}%
  \BibitemOpen
  \bibfield  {author} {\bibinfo {author} {\bibfnamefont {Niculescu}\
  \bibnamefont {V}}, \bibinfo {author} {\bibfnamefont {T.~J.}\ \bibnamefont
  {Burch}}, \bibinfo {author} {\bibfnamefont {K.}~\bibnamefont {Rag}}, \ and\
  \bibinfo {author} {\bibfnamefont {J~.}\ \bibnamefont {Budnick}},\ }\bibfield
  {title} {RMX Compounds Fromed by Alkaline Earths, Europium and ytterbium - I.
  Ternary Phases with M = Cu, Ag, Au; X= Sb, Bi,\ }\href@noop {} {\bibfield
  {journal} {\bibinfo  {journal} {J. Less Common Met.}\ }\textbf {\bibinfo
  {volume} {166}},\ \bibinfo {pages} {319--327} (\bibinfo {year}
  {1990})}\BibitemShut {NoStop}%
\end{thebibliography}%

\end{document}